\documentclass[12pt]{iopart} 
\usepackage{graphicx,epsfig}

\def\ttb{$t \overline{t}$ }
\def\ttbb{$t \overline{t}b \overline{b}$ }

\def\pT{$P_{T}$ }
\def\emiss{$\not\hskip-5truedd E_{T}$ }
 
\begin{document} 
\def\beq{\begin{eqnarray}}    %%%  begequation/eqnarray 
\def\eeq{\end{eqnarray}}      %%%  endequation/eqnarray 
\def\be{\begin{equation}}   
\def\eq{\end{equation}}      
\def\mg{m_{\tilde{g}}} 
\def\msb1{m_{\tilde{b}_{1}}} 
\def\mst1{m_{\tilde{t}_{1}}} 
\def\rpv{\mbox{$\rlap{\kern0.25em/}R_p$ }}
\def\nn{\nonumber}

\begin{flushright}
{FSU-HEP-031231}
\end{flushright}

\title{Signals~from~R-parity~violating~top~quark~decays~at~LHC}

\author{A. Belyaev$\dag$, M-H.~Genest$\ddag$, C.~Leroy$\ddag$, 
R.~Mehdiyev$\ddag$
\footnote[7]{On leave of absence from Institute
of Physics, ANAS, Baku, AZ-370143, Azerbaijan
}}

\address{\dag\ Physics Department, Florida State University,\\ Tallahassee, FL
32306-4350, USA.}

\address{\ddag\ Universit\'e de Montr\'eal, D\'epartement de Physique,\\
		  Montr\'eal, QC, H3C 3J7, Canada.}

\begin{abstract}  We evaluate the potential of the CERN LHC collider to
observe rare decays of the top quark in channels involving R-parity violating 
(\rpv) interactions.
We stress the importance of calculating top quark  production and
decay simultaneously as a true $2\to 4$ process. \\
The process of $t\bar{t}$ pair production followed by \rpv decay of one of 
the top quarks is analyzed with fast detector simulation.  We show that intermediate
supersymmetric particles can be observed as resonances even if they are heavier
than the top quark due to the  significant off-shell top-quark mass effects.
 The  approach where the top quark is produced 
on-mass-shell and then decays into 2- or 3-body  final state would in general
lead  to incorrect kinematical distributions and rates.
The rates of the  $2\to 4$ process with   top quark production and  \rpv 3-body
decay  depend on the total width of the  heavy intermediate sfermion which could,
therefore, be measured indirectly.\\
We find that the LHC collider offers a unique potential to study rare top quark
decays in the framework of supersymmetry with broken $R$-parity for  branching fractions
of \rpv top decays as low as  $\simeq 10^{-6}$.

\end{abstract}



\maketitle

%%%%%%%%%%%%%%%%%%%%%%%%%%%%%%%%%%%%%%%%%%%%%%%%%%%%%%%%%%%% 
 
%%%%%%%%%%%  INTRODUCTION   %%%%%%%%%%%%%%%%%%%%%%%%%%%%%%%%% 
 
\section{Introduction}
 
Processes involving the top quark, the heaviest known fermion with a  mass
close to the scale of the  electroweak symmetry breaking, offer a unique
possibility to search for physics beyond the Standard Model(SM). It is well known 
that the SM suffers from fundamental problems such as hierarchy, fine-tuning, 
and absence of dark matter candidates. It is therefore believed that there must
exist a more fundamental underlying theory.
One of the most promising alternative models which resolves these problems 
is supersymmetry. 
The Minimal Supersymmetric extension of the Standard
Model(MSSM), with the gauge group $SU(3)_c\times SU(2)_L\times U(1)_Y$
contains  SM particles, their superpartners and an additional Higgs
doublet. The MSSM possesses an additional discrete  symmetry, called
$R$-parity ($R_p$ ), which conserves lepton and baryon number:

\begin{equation}
\hspace{3truecm}
R_p=(-1)^{3B+L+2S}, 
\end{equation} 
where $B$, $L$ and $S$ are the baryon number, lepton number and spin,
respectively.  However, there is no clear theoretical preference 
for the status of  $R_p$ -- to be conserved or violated -- in the 
supersymmetric theories. Therefore the answer whether the parity is realized 
in nature or not should be given by experiment.
It is important to notice that \rpv\ is much less constrained in general
for the 3-rd fermion-sfermion generation compared to the first  two generations.
Therefore the CERN LHC collider, which can be considered as a top quark factory,
plays a special role as a the perfect tool for testing the conservation of $R_p$
in the top-stop sector.
With $t\bar{t}$ production cross section of the order of 
$800~pb$~\cite{Bonciani:1998vc}, $\sim 10^8$ top quarks will be produced per year,
assuming an integrated luminosity of 100~$fb^{-1}$. These high statistics will
allow for precision measurements of top quark physics
and in particular, for high sensitivity to
rare top quark decays in a \rpv scenario.

Rare top-decays with $\rpv$ have been intensively studied for the last
decade, and even before the discovery of the top quark~\cite{Dreiner:1991dt}. 
If sfermions are light enough, the 2-body  decay of top quark into fermion-sfermion pair
will take place
~\cite{Agashe:1996qm,Bartl:1996gz,deCampos:1998kw,Bar-Shalom:1998uq,
Navarro:1999tz,Han:1999qs,Abraham:2000kx,Abraham:2001rj}, otherwise
the decay with a single neutralino (${\chi}^0$)
(hereafter we are using  ${\chi}^0$ notation for the lightest neutralino) 
in the final
state proceeds with the kinematics of a 3-body decay~\cite{Han:1999qs,Belyaev:2000pg}. 

The aim of the present paper is to take a closer look at
\rpv 3-body top quark decay expanding on previous studies
in the following way:
\begin{itemize}
\item we treat top quark production and decay
coherently (i.e. calculate diagrams with off-mass-shell top quark)
and show the crucial role of this approach for the predictions
of kinematical distributions and  production rates;
\item we perform a simulation of the processes at detector level,
thus obtaining realistic predictions for the reach at the LHC.
\end{itemize}

The outline of the paper is as follows: In Sect.~2, we evaluate production
rates for the  process $pp\to t \tilde{\chi}^0 q\ell$. 
We show that only by calculating together top quark
production and decay do we obtain the correct cross section.
We scan the $\mu-M_2$  
plane and present parton-level kinematical distributions.  
Sect.~3 gives details of the fast Monte-Carlo simulation, including the effect
of initial and final state radiation, as well as hadronization effects, followed
by the fast detector simulation.
In  Sect.~4, we describe the analysis procedure for
extracting the signal from backgrounds and reconstructing the top quark 
and the slepton.  
In Sect.~5, we estimate
the sensitivity of the LHC for observing of \rpv top quark decays
and  we draw our conclusions in Sect.~6.

\section{Top quark production and decay: \rpv scenario}

The $\rpv$ superpotential of the MSSM  can be written 
as~\cite{Weinberg:1982wj,Sakai:1982pk,Dimopoulos:1982dw}
\begin{equation} \label{rpvpot} 
{\cal W}_{\rpv}=\lambda_{ijk}L_iL_j\bar{E_k}
+\lambda_{ijk}'L_iQ_j\bar{D_k}
+\lambda^{\prime\prime}_{ijk}U_{i}^cD_{j}^c\bar{D_{k}}
+\mu_iL_iH_2,
\end{equation}
where $L_i(Q_i)$ and $E_i(U_i,D_i)$ are the left-handed lepton (quark) 
doublet and right-handed lepton (quark) singlet chiral superfields; 
$H_2$ is the second Higgs doublet field; $\mu_i$ are  bi-linear \rpv couplings;
$i,j,k$ are generation indices. 
Terms with $\lambda$ and $\lambda^\prime$ coefficients
violate lepton number while the term with coefficient $\lambda^{\prime\prime}$
violates baryon number.
Lepton- and baryon- number violating 
operators together would lead to fast proton decay.
It has been shown~\cite{Ibanez:1992pr,Lola:1993ip,Ellis:1998rj}
that there exist symmetries that allow the \rpv\ for the subset 
of these operators consistent with the limits on proton decay.
Therefore, here we  assume hereafter that the baryon-number violating operator
is vanishing. The alternative case, with non-vanishing baryon-number 
violating operator but no lepton-number violating operator,
leads to hadronic top quark decay, which is very difficult
to observe due to the huge QCD background. 

The last bi-linear $R$-parity violating (BRPV) 
term in the superpotential~(\ref{rpvpot}) is responsible for 
spontaneous \rpv violation. In general, it cannot be rotated away 
from the superpotential together with the SUSY soft breaking 
term by a suitable choice of the basis~\cite{Diaz:1998xc}.
If one tried to do this, then \rpv\ terms would be reintroduced via two
trilinear terms given by $LL\bar{E}$ and $LQ\bar{D}$ 
in equation~(\ref{rpvpot}).
Since our specific study of \rpv\ involves only the top quark decay,
the $LQ\bar{D}$ \rpv\ operator (precisely, $LQ_3\bar{D}$) is given
by the second term of~(\ref{rpvpot}), 
which also can effectively arise from BRPV term. 
It leads
to the following Lagrangian in four-component Dirac notation:

%%%%%%%%%%%%%%%%%%%%%%%%%%%%%%%%%%%%%%%%%%%%%%%%%%%%%%%%%%%%%%%%%%%% 
\begin{eqnarray} \label{lqdbar}
L_{\rpv}&=&\lambda'_{ijk}[\tilde{\nu}_{iL}\bar{d}_{kR}d_{jL}+
\tilde{d}_{jL}\bar{d}_{kR}\nu_{iL}+
{\tilde{d}^*}_{kL}\overline{(\nu_{iL})^C}d_{jL} - \nn \\
&&\tilde{e}_{iL}\bar{d}_{kR}u_{jL}-\tilde{u}_{jL}\bar{d}_{kR}e_{iL}-
{\tilde{d}^*}_{kL}\overline{(e_{iL})^C}u_{jL}]
 + h.c.
\end{eqnarray}
%%%%%%%%%%%%%%%%%%%%%%%%%%%%%%%%%%%%%%%%%%%%%%%%%%%%%%%%%%%%%%%%%%%% 

\begin{figure}[htb] 
\centerline{\epsfig{file=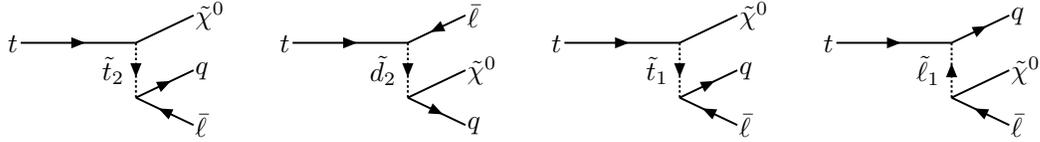,width=\textwidth}}
\caption{Tree-level diagrams for three-body top quark decay
into neutralino ($\tilde{\chi}^0$), lepton (l) and quark (q) via \rpv operators. 
\label{diag-top-decay}
}
\end{figure}

Since we study \rpv in the third family, only  $\lambda'_{i3k}$ couplings are relevant.
They give rise to top quark decay into a single
neutralino  via  the Feynman  diagrams shown in 
Fig.~\ref{diag-top-decay}.
%%%%%%%%%%%%%%%%%%%%%%%%%%%%%%%%%%%%%%%%%%%%%%%%%%%%%%%%%%%%%%%%%%%% Fig.1
%%%%%%%%%%%%%%%%%%%%%%%%%%%%%%%%%%%%%%%%%%%%%%%%%%%%%%%%%%%%%%%%%%% 
Three kinds of virtual fermions can be present in the 
diagrams for non-vanishing $\lambda'_{i3k}$: 
a slepton, a squark of the first or second
generation and a $stop$-quark.

The decay width for this process of top-quark decay depends on:

$\bullet$ The magnitudes of the
neutralino and chargino masses and couplings,
which are functions of the MSSM
parameters. Assuming gaugino-mass unification,
the $U(1)$ and $SU(2)$ gaugino masses $M_1$ and
$M_2$  are related by $M_1 = \frac{5}{3}\tan\theta_W^2 M_2$,
in which case the gaugino and higgsino masses and couplings
are determined by $M_2$ -- $SU(2)$ gaugino mass, 
$\mu$ -- superpotential Higgs mass parameter
and $\tan\beta$.

$\bullet$ The masses of the intermediate sfermion -- squarks and sleptons --
which could be either lighter or heavier than the top-quark.

\begin{figure}[htb] 
\epsfig{file=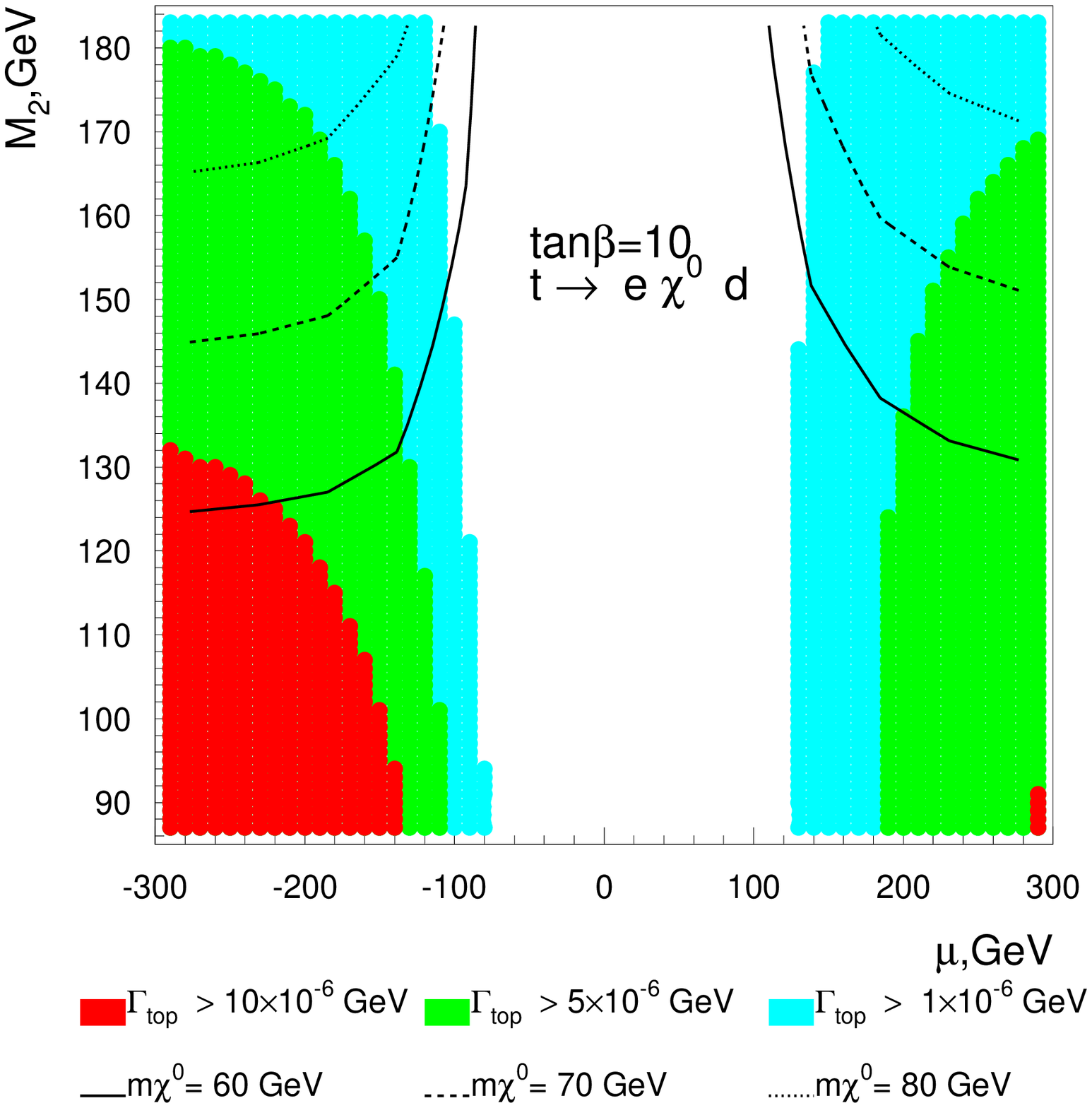,width=0.5\textwidth}
\epsfig{file=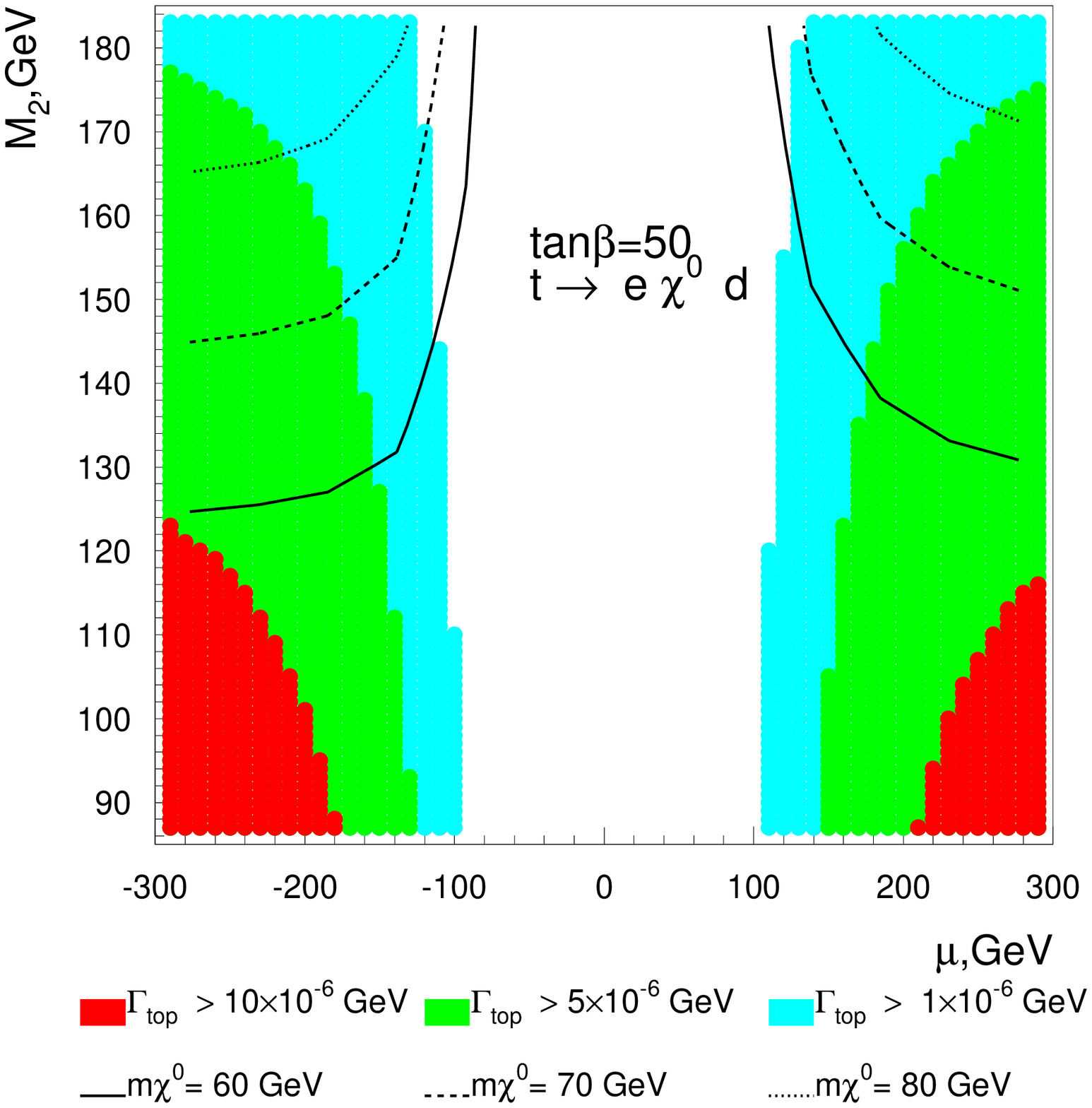,width=0.5\textwidth}
\caption{Contours for the partial 3-body top decay width
$\Gamma_{top}[t\to\tilde{\chi}^0 \ell q]$\ \label{3-body-decay}
for a slepton mass $m_{\tilde{l}}=200~GeV$}
\end{figure}

We assume here for simplicity
the case where one sfermion, for example, a slepton, is lighter
than any other intermediate sfermions, i.e. it gives the leading contribution to the
\rpv decay of the top quark.
For our calculations we have used CompHEP v33.23~\cite{Pukhov:1999gg}
together with a fully implemented model of \rpv~(Eq.\ref{rpvpot})
interactions~\cite{ABAG-rpvmodel}.
Contours of the partial width, $\Gamma_{top}$, for top decay  by the channel 
$t\to\tilde{\chi}^0 \ell q$
in the $(\mu-M_2)$  plane 
 are shown in Fig.~\ref{3-body-decay}.
The partial width was calculated as in~\cite{Belyaev:2000pg}.

%%%%%%%%%%%%%%%%%%%%%%%%%%%%%%%%%%%%%%%%%%%%%%%%%%%%%%%%%%%%%%%%%%%% Fig.2
%%%%%%%%%%%%%%%%%%%%%%%%%%%%%%%%%%%%%%%%%%%%%%%%%%%%%%%%%%%%%%%%%%%% 

\begin{figure}[htb] 
\epsfig{file=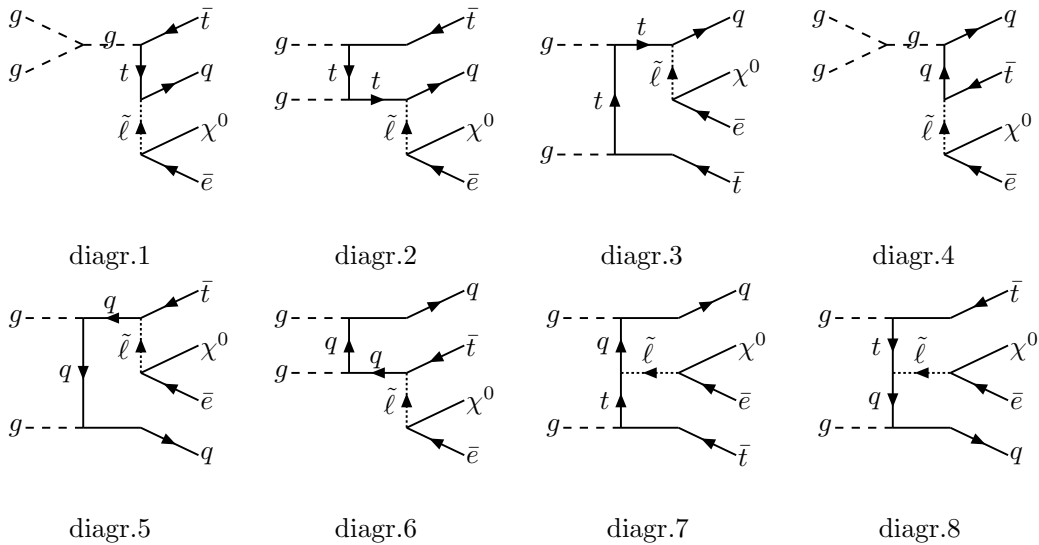,width=\textwidth}
\caption{Tree-level diagrams for the $gg\to t \tilde{\chi}^0 \ell q$
process of $t\bar{t}$ production and decay due to \rpv
interactions\label{2-4-diag}}
\end{figure}

The decay width decreases as $M_2$ increases since the neutralino mass
increases with this parameter.
The cross section of the complete $2\to 4$
process $gg\to t \tilde{\chi}^0 \ell q$ describing $t\bar{t}$ production
and \rpv decay of one of top quarks can now be evaluated.
The corresponding diagrams are shown in Fig.~\ref{2-4-diag}.
%%%%%%%%%%%%%%%%%%%%%%%%%%%%%%%%%%%%%%%%%%%%%%%%%%%%%%%%%%%%%%%%%%%% Fig.3
%%%%%%%%%%%%%%%%%%%%%%%%%%%%%%%%%%%%%%%%%%%%%%%%%%%%%%%%%%%%%%%%%%% 

For the parton-level calculations, we used the CTEQ5L~\cite{Lai:1999wy} 
parton density function with
the top quark mass as the QCD scale.
In order to avoid soft and collinear divergences when massless 
quarks are present in the final
state, we applied a quark transverse momentum cut, $p^q_T>10$~GeV. 
Diagrams (1-3) in Fig.~\ref{2-4-diag}
form a gauge-invariant subset which does not require the $p^q_T$ cut.
However, the total set of subprocesses leading to $t \tilde{\chi}^0 \ell q$
state consists of 8 Feynman diagrams which should be calculated together.

\begin{figure}[htb] 
\epsfig{file=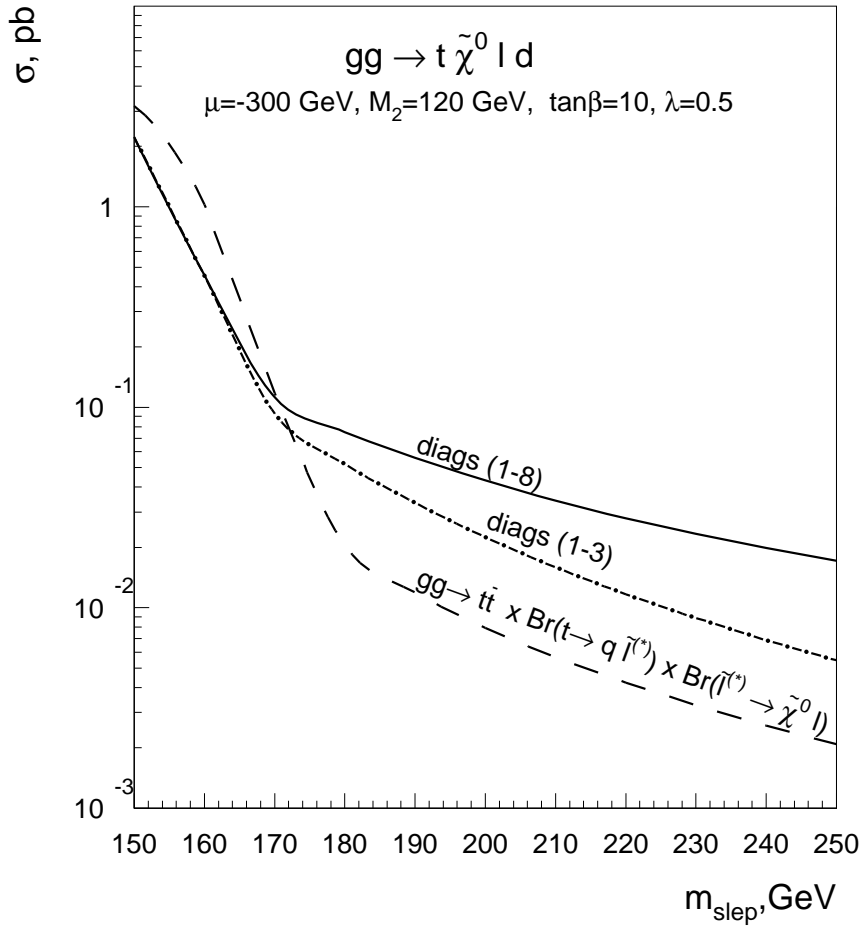,height=0.85\textwidth}
\centering{
\caption{Cross section of $gg\to t \tilde{\chi}^0 \ell q$
process as a function of the slepton mass for the 3 approaches (see text).
\label{cs23}}
}
\end{figure}

In Fig.~\ref{cs23}, we plot the cross section for the process
$gg\to t \tilde{\chi}^0 \ell q$ as a function of the slepton mass 
under three approaches:

(i) $gg\to t \bar{t}\times Br[t\to q \tilde\ell]\times Br[\tilde\ell\to \ell \chi^0]$
    ($m_{\tilde\ell}< m_{top}$)
     or $gg\to t \bar{t}\times Br[t\to q \ell \chi^0]$
     ($m_{\tilde\ell}> m_{top}$). Here, the top quark undergoing \rpv decay 
      stays on-shell

(ii) contribution of diagrams (1-3) only 

(iii) complete calculation of the process
    $gg\to t \tilde{\chi}^0 \ell q$, i.e. calculations of the set of 
    diagrams (1-8).

\begin{figure}[htb] 
\epsfig{file=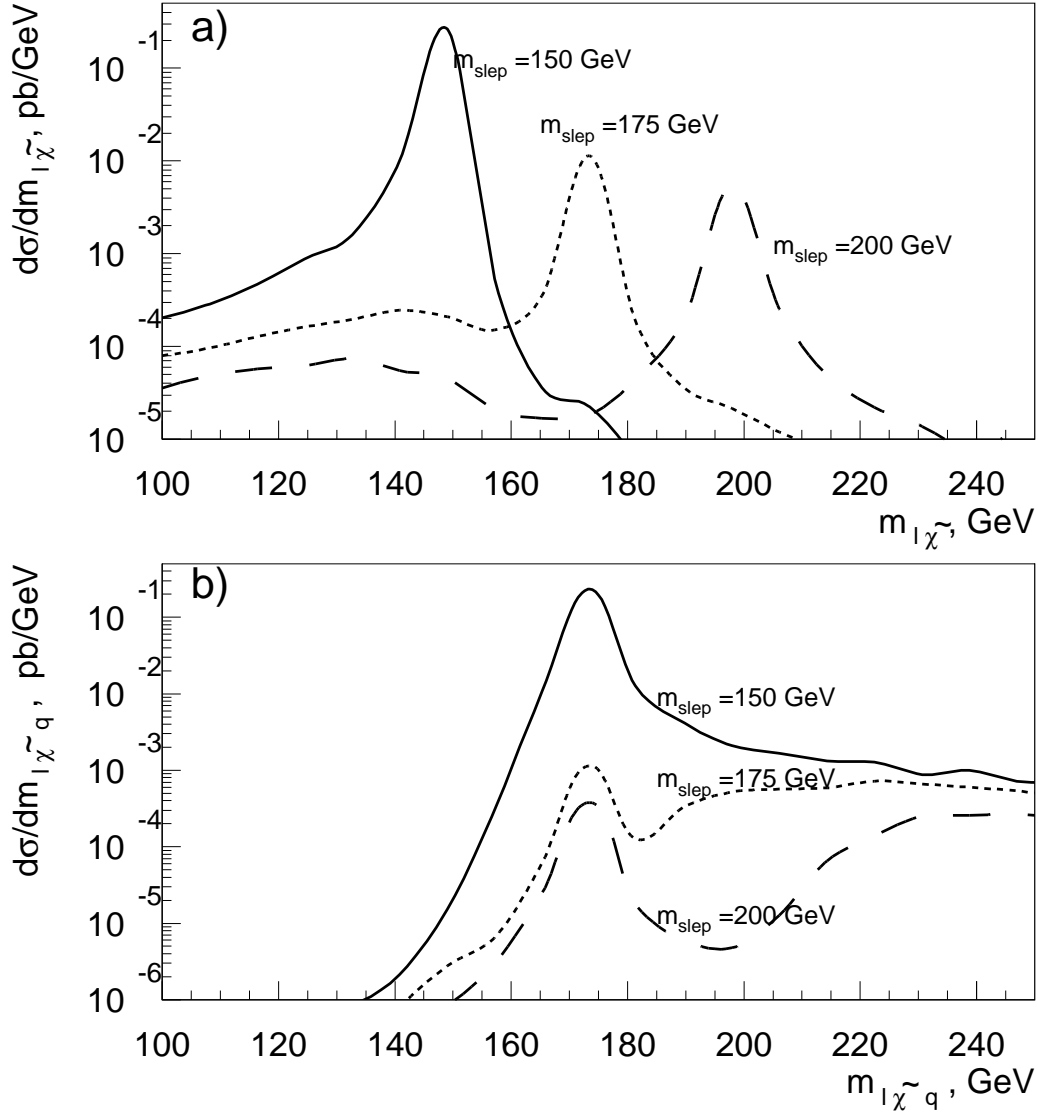,width=0.95\textwidth}
\centering{
\caption{Parton-level $\tilde{\chi}^0 \ell$  invariant  mass distribution (top)
and $\tilde{\chi}^0 \ell q$  invariant mass distribution(bottom) for different 
slepton masses.
\label{2-4-dist}}
}
\end{figure}

One can see that the cross section evaluated in approach (i) agrees within 30-50\%
with those of  (ii) or (iii) when $m_{\tilde\ell}< m_{top}$.
On the other hand, when  $m_{\tilde\ell}> m_{top}$ approach (i)
underestimates the true cross section of approach (iii)
by about an order of magnitude!
Rates predicted by approach (ii) are about a 
factor of 2.5 smaller compared with those of approach (iii).
To understand the origin of the difference between approaches (i) and (iii),
we studied kinematical distributions. 
Fig.~\ref{2-4-dist} shows the $\tilde{\chi}^0 \ell$ (top)
and  $\tilde{\chi}^0 \ell q$ (bottom) 
invariant  mass distributions for the complete calculation of (iii).

%%%%%%%%%%%%%%%%%%%%%%%%%%%%%%%%%%%%%%%%%%%%%%%%%%%%%%%%%%%%%%%%%%%% Fig.4
%%%%%%%%%%%%%%%%%%%%%%%%%%%%%%%%%%%%%%%%%%%%%%%%%%%%%%%%%%%%%%%%%%%% 
%
Even for a quite heavy slepton ( $m_{\tilde\ell} = 200$~GeV ), the resonant
peak from the slepton in  $\tilde{\chi}^0 \ell$ distribution is clearly seen. 
This means that the slepton forces the top quark  to be off-shell and becomes the
resonance itself. The contribution from the on-shell heavy slepton is
significant. 
In approach ii.) the $\tilde{\chi}^0 \ell$ distribution
is similar, but the  slepton resonance peak is slightly suppressed.

%%%%%%%%%%%%%%%%%%%%%%%%%%%%%%%%%%%%%%%%%%%%%%%%%%%%%%%%%%%%%%%%%%%% Fig.5
%%%%%%%%%%%%%%%%%%%%%%%%%%%%%%%%%%%%%%%%%%%%%%%%%%%%%%%%%%%%%%%%%%%% %

The $q \tilde{\chi}^0 \ell$
invariant mass  distribution  (bottom of Fig.~\ref{2-4-dist}) 
in the top quark decay is sensitive to the mass of the intermediate slepton.
For a heavy slepton $m_{\tilde\ell} \ge m_{top}$, 
the top quark resonance peak is strongly suppressed and 
a right shoulder appears in the
$\ell\tilde{\chi}^0 q$ invariant mass spectrum. This happens because the heavy
slepton "pushes" the top quark to be off-shell and  becomes itself on-shell. 
In this case, the experimental reconstruction of the top quark mass 
is more difficult due to this shoulder.
On the contrary,  if the slepton mass is less than the top quark mass,  two-body
top quark decay takes place and one can observe a pronounced peak in the $\ell
\tilde{\chi}^0 q$ invariant mass distribution.

We have explicitly checked that total cross sections 
for approaches (i), (ii) and (iii) agree fairly well 
(at about 20\% level)
after the application of a 20 GeV mass window cut on 
$\ell\tilde{\chi}^0$ invariant mass to remove
the contribution from the  slepton.  

\begin{figure}[htb] 
\epsfig{file=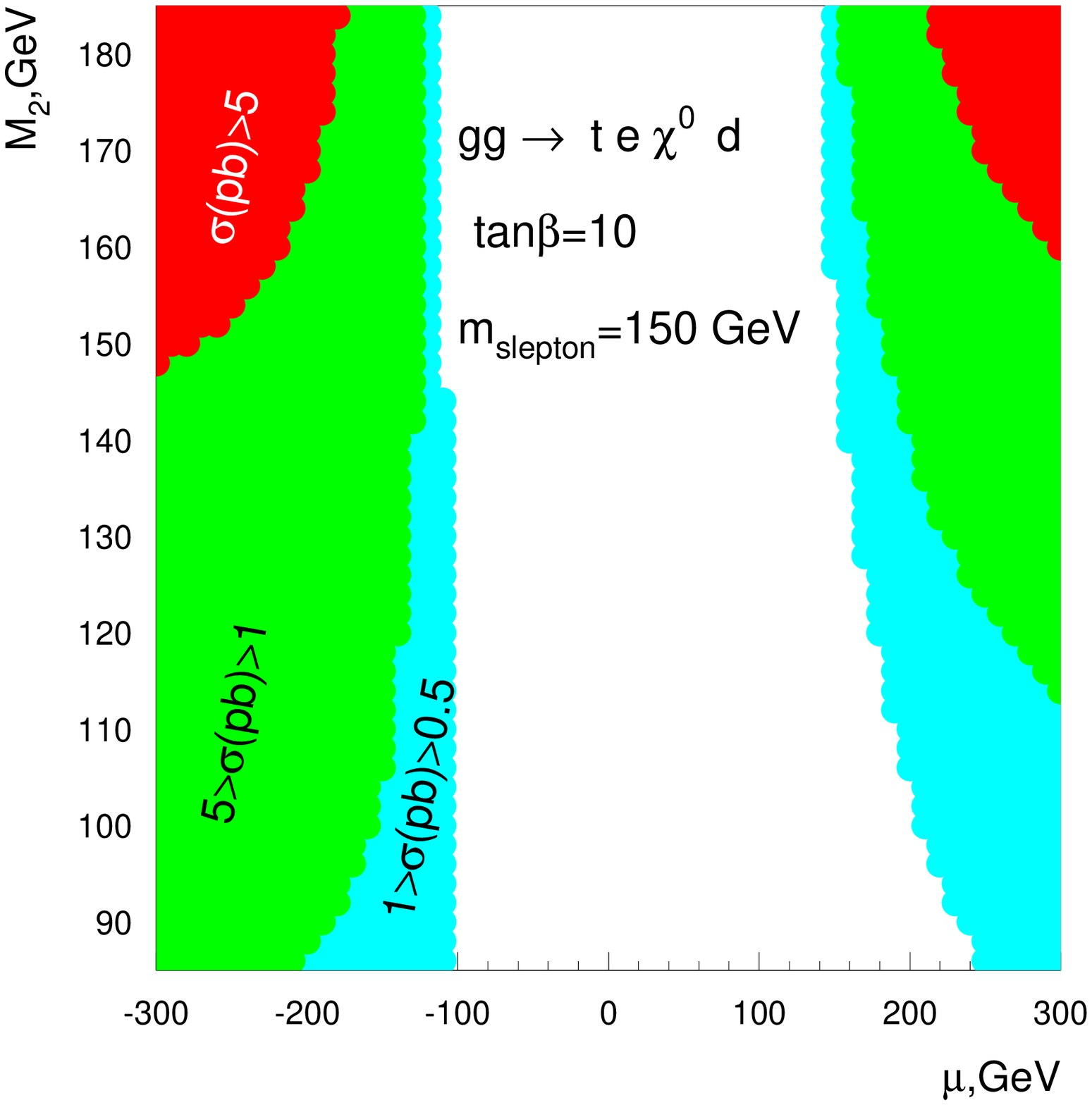,width=0.5\textwidth}
\epsfig{file=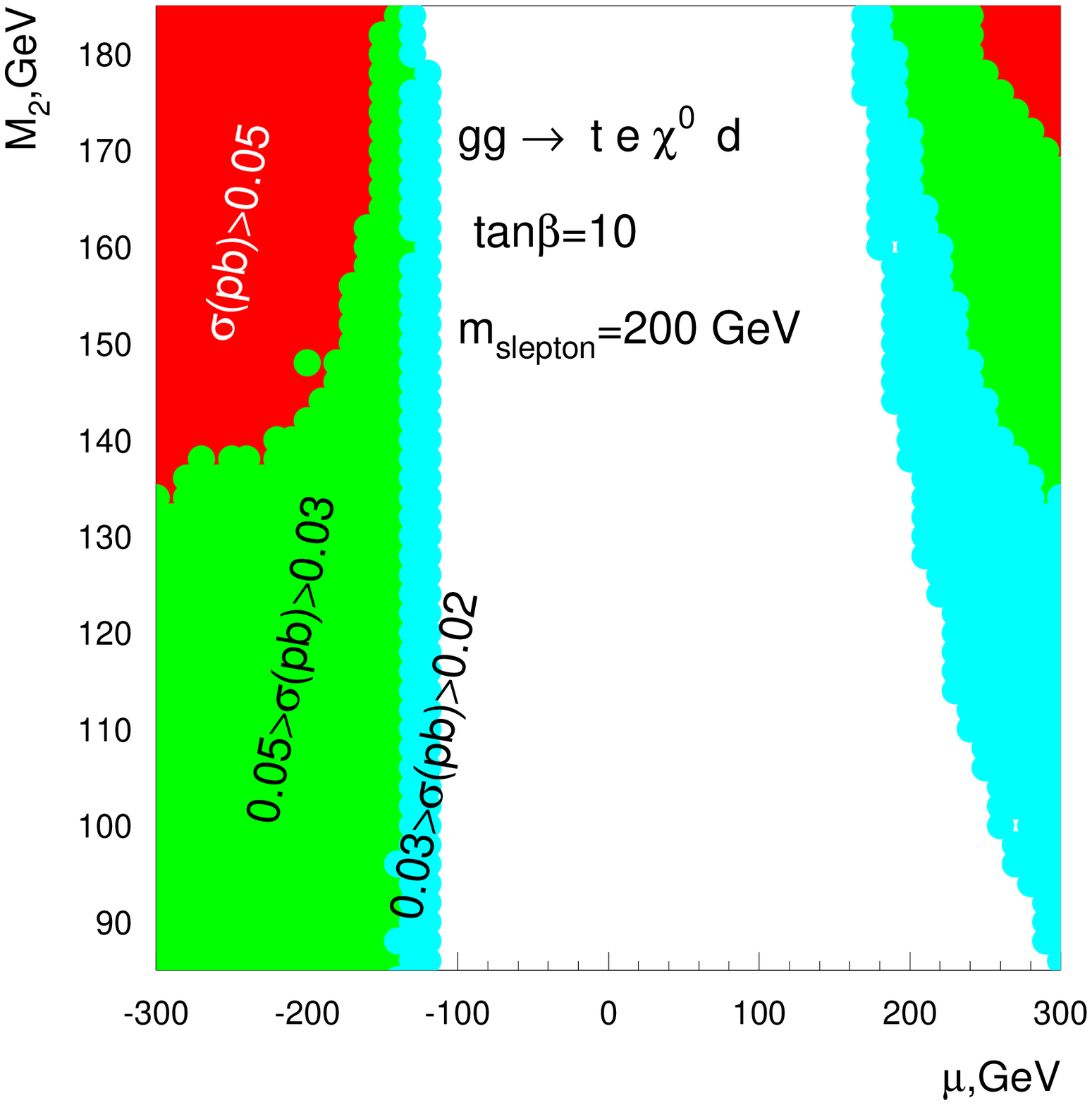,width=0.5\textwidth}
\caption{Contours for the cross-section of $gg\to t \tilde{\chi}^0 \ell q$
process for $m_{\tilde\ell}=150$~GeV (left) and  $m_{\tilde\ell}=200$~GeV 
(right).\label{cs-mu-m2}}
\end{figure}

Another difference between the predictions of approaches (i) and
(iii) must be pointed out.  In Fig.~\ref{3-body-decay}, one can see that the 
$\rpv$ top decay width decreases with an increase of the $M_2$ parameter. 
Therefore, in approach (i), the cross section for the 
$gg\to t\tilde{\chi}^0  \ell q$ process
also decreases as $M_2$ increases. The situation is just the opposite in 
approach (iii), because the cross section of the $true$  $2\to 3$ process also
depends on the {\it slepton width}. In Fig.~\ref{cs-mu-m2}, one can see that the
cross section increases with $M_2$ due a decrease of the slepton width. 

For the rest of the analysis, we use approach (iii) which gives the correct
cross section and kinematical distributions. 

One should also note that we treated neutralino as an on-shell particle.
It decays through the intermediate sparticle, which 
could also, in principle, "push" a neutralino to be off-shell
and invalidate the on-shell approach for the neutralino.
However one should recall that
neutralino width is at least several orders of magnitude
smaller than the width of the intermediate scalar
($\sim 10^{-6}-10^{-8}$ GeV for $\lambda\sim 0.1-0.2$ and 
$m_{\tilde{q},\tilde{l}} \sim 100-200$ GeV).
\footnote{In principle, the neutralino width could be comparable with the
 width of a slepton, when the $\chi^0_1\to  \tilde{t}t$ 2-body decay channel 
becomes open if $m_{\tilde{l}} > m_\chi^0 > m_t $.
But such case is not relevant to our study since in this scenario the
RPV top-quark decay branching ratio will be
 several orders ($\sim 3-4$) of magnitude below the sensitivity of the LHC.}
This fact eliminates the
probability  for the intermediate sparticle to be on-shell for 
the $\tilde{\chi^0}\to  \ell^+(\tilde{\ell^{-}_{*}} ) \to d q$ chain.
The asteriks here denotes an off-shell intermediate sparticle. 
We have checked
this qualitative argument numerically and have  calculated the width and mass 
distributions for the $t\to d (\tilde{\ell^{+}}) \to \ell^{+} 
({\chi^0}) \to \ell^+ (\tilde{\ell^{-}_{*}}) \to d q$
process for true $1\to 5$ top quark decay.
We have chosen $m_{\tilde{l}}$=150 GeV and have found that,
indeed, the invariant mass distribution of  the $l d q$ system
from $\tilde{\chi^0}\to  \ell^+(\tilde{\ell^{-}_{*}} ) \to d q$ process,  
has a delta-function-like shape and the exact width for the 
$1\to 5$ top quark decay
can be accurately (within numerical errors of ~1\%) reproduced
by  product of the $1\to 3$ top quark decay width and the neutralino branching ratio.
This validates the treating of neutralino as on-shell particle
and suffices dealing with 3-body top decay.

In the  next section, we
perform a fast Monte Carlo simulation of the $gg\to t \tilde{\chi}^0 \ell q$
process in order to estimate a realistic experimental sensitivity to \rpv
top quark decay.

%%%%%%%%%%%%%%%%%%%%%%%%%%%%%%%%%%%%%%%%%%%%%%%%%%%%%%%%%%%%%%%%%%%% Fig.6
%%%%%%%%%%%%%%%%%%%%%%%%%%%%%%%%%%%%%%%%%%%%%%%%%%%%%%%%%%%%%%%%%%%% 

\section{Signatures and details of simulation}

We have used the CompHEP package  to generate \ttb events with 
\rpv decay of one of the top-quarks --- $t\to \tilde{\chi}^0 q\ell$
and SM decay of the other one --- $\overline{t}\to Wb$.
We have studied the cases of hadronic and leptonic
decay modes of the $W$-boson from the top-quark undergoing the SM decay.
In case of the leptonic decay of $W$,
we  studied  $W \rightarrow \mu\nu$ decay channel  in order to 
enhance the separation between the top \rpv SUSY and SM decay chains.

In the present study we assume that
{\it only one $\lambda'_{i3k}$ parameter in Eq. (3) is non-vanishing}.
A value of 0.5 for $\lambda'_{131}$  was chosen for a reference. 
The following values of supersymmetric
parameters were used:  a gaugino mass $M_2=120$~GeV,
a higgsino mass parameter $\mu=-300$~GeV, $tan\beta=10$, and a
slepton mass either $m_{\tilde\ell}=150$~GeV ($m_{\tilde\ell}< m_{top}$) or
$200$ GeV ($m_{\tilde\ell}> m_{top}$). 
Under these conditions, the total cross-sections obtained with CompHEP 
for the $2\to 4$ signal
process (see Fig.~\ref{2-4-diag}, $p_T^q>10$~GeV) are 8~pb 
for a slepton mass $m_{\tilde\ell}=150$~GeV and 480~fb 
for $m_{\tilde\ell}=200$~GeV. This is to be compared with
the total cross section of 833~pb~\cite{Bonciani:1998vc}
for the \ttb background.

\begin{figure}[htb] 
\vspace*{-0.5cm}
\epsfig{file=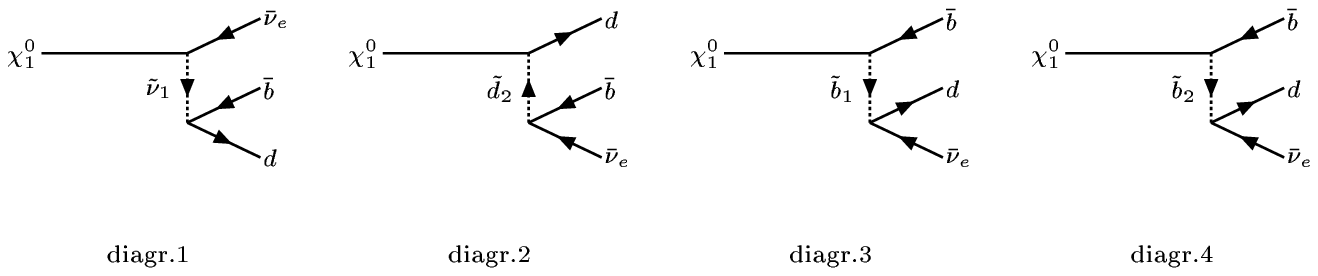,width=\textwidth}
\epsfig{file=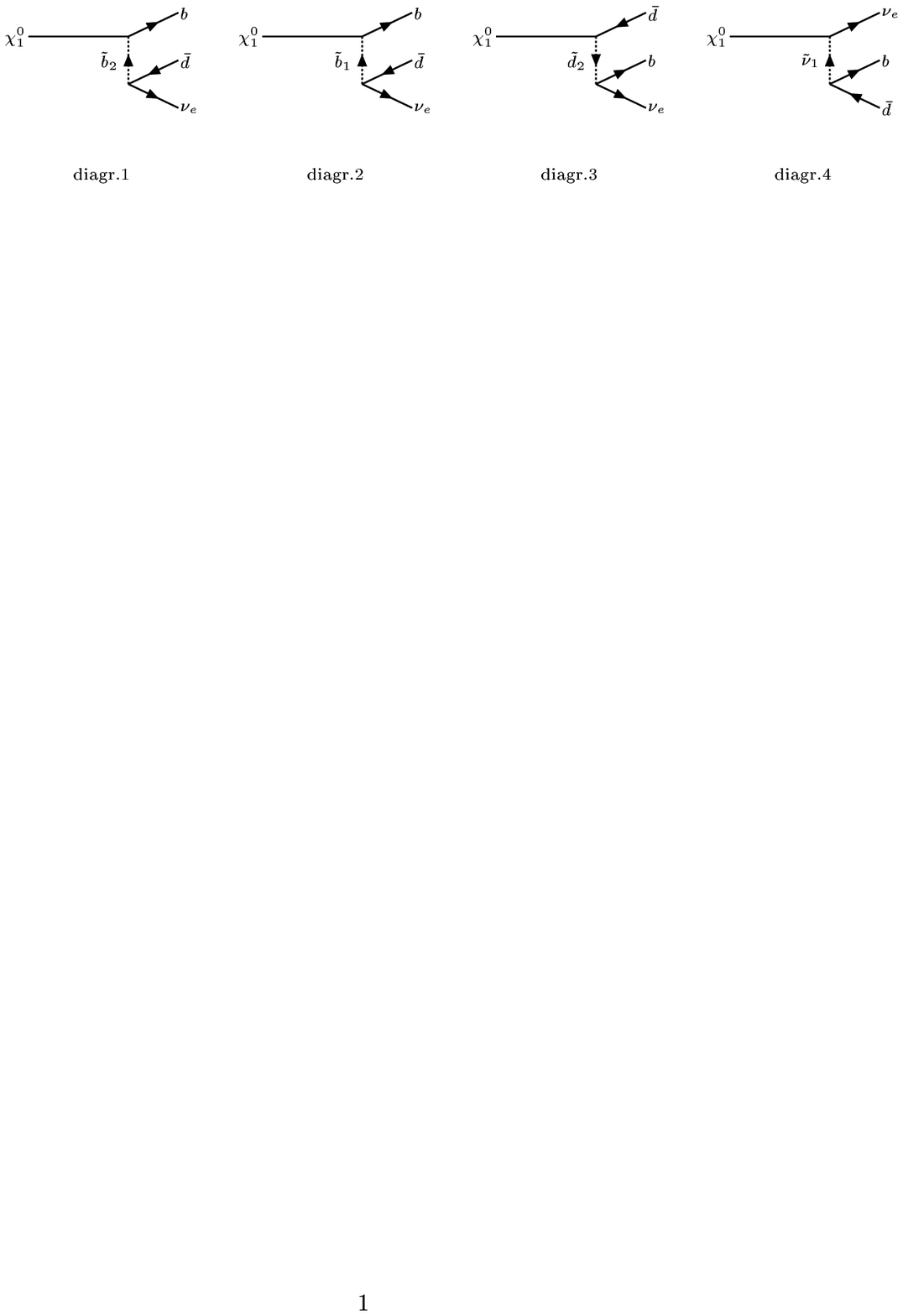,width=\textwidth}
\caption{Tree-level diagrams for MSSM 
$\tilde{\chi}^0\to \bar{\nu}_e\bar{b}d(\nu_e b\bar{d})$ decay. 
\label{o1_mssm}}
\end{figure}

With this set of parameters, the
mass of the neutralino is $m_{\chi^0}=58$~GeV.  If 
$\lambda'_{i3k}$ is non-zero only for $i=k=1$, the neutralino  will decay into
$\bar{\nu}_e\bar{b}d$ or $\nu_e b\bar{d}$ ($BDN$) (Fig.~\ref{o1_mssm}). 

\begin{figure}[htb] 
\epsfig{file=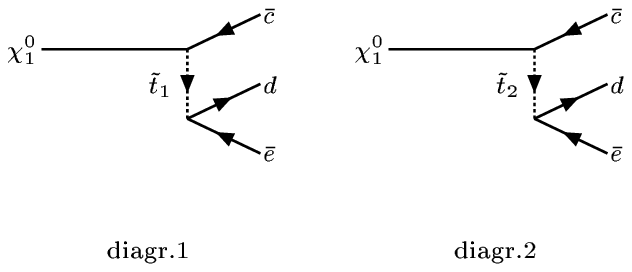,width=0.5\textwidth}
\epsfig{file=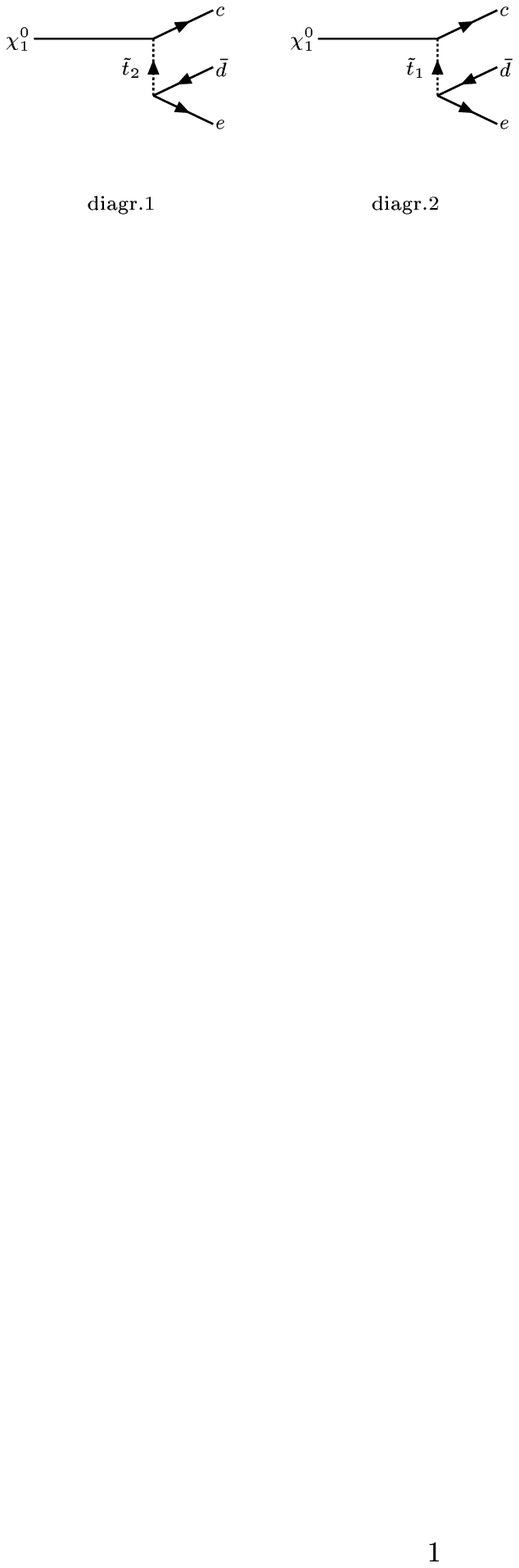,width=0.5\textwidth}
\caption{Tree-level diagrams for FCNC $\tilde{\chi}^0\to 
         \bar{c}d\bar{e}(c\bar{d}e)$ decay.\label{o1_fcnc}}
\end{figure}

We also consider the possibility of
stop-scharm mixing leading to neutralino decay into   $\ell^- c\bar{d}$
or $\ell^+ \bar{c}d$ (Fig.~\ref{o1_fcnc}).  It must be pointed out that
such mixing, leading to flavor changing  neutral currents (FCNC),
is practically not constrained by experimental
data~\cite{Misiak:1998ei,Diaz-Cruz:2001gf}.  We assume maximum 
stop-scharm mixing here, but the results can be easily rescaled for any
arbitrary value of the mixing. If the stop-scharm mixing is
significant, and if one of the stop quarks is lighter than other
sfermions appearing in the Feynman diagrams, then the
$\tilde{\chi}^0\to \bar{c}d\bar{e}(c\bar{d}e)$ ($CDE$) channel could be
dominant.  In this case, one would  be able  to obtain with much higher efficiency 
( in comparison to $BDN$ case ) a narrow peak for the reconstructed top quark mass.

An additional advantage of this decay mode is that the
neutralino, by virtue of its Majorana nature,
decays equally to positively and negatively charged leptons.
This means that half of all events would contain two
like-sign leptons.  This clean signal signature allows the application of an
effective cut to suppress possible backgrounds. The
study of the \rpv rare top decays at the LHC would also allow the direct 
measurement of, or the establishment  of a limit on stop-scharm mixing.
Below, we perform an analysis under both assumptions for neutralino decay: 
with and without stop-scharm mixing.

%%%%%%%%%%%%%%%%%%%%%%%%%%%%%%%%%%%%%%%%%%%%%%%%%%%%%%%%%%%%%%%%%%%% Fig.7  
%%%%%%%%%%%%%%%%%%%%%%%%%%%%%%%%%%%%%%%%%%%%%%%%%%%%%%%%%%%%%%%%%%%% 
%%%%%%%%%%%%%%%%%%%%%%%%%%%%%%%%%%%%%%%%%%%%%%%%%%%%%%%%%%%%%%%%%%%% Fig.8
%%%%%%%%%%%%%%%%%%%%%%%%%%%%%%%%%%%%%%%%%%%%%%%%%%%%%%%%%%%%%%%%%%%% 

PYTHIA 6.2 \cite{pythia:man} was used to account for initial
and final state radiation and to perform hadronization and decay
of resonances. The Lund symmetric fragmentation function has been used for 
light flavors, but charm and heavier ones - according to the Peterson/SLAC 
function. 
The simulation of the signal and background was performed with ATLFAST
\cite{atlfast:man} to take into account the  experimental conditions
prevailing at LHC for the ATLAS detector\cite{atlas:tp}. 

Jet energies were reconstructed by clustering
hadronic calorimeters cells
of  dimensions $ \Delta \eta \times \Delta \phi =0.1\times 0.1 $
within the pseudorapidity range $ -2.5<\eta <2.5 $, where $\phi$ is the
azimuthal angle.  The hadronic energy resolution of the ATLAS detector is 
parametrized as $0.5/\sqrt{E(GeV)}{\oplus }0.03 $  over this $ \eta  $ region. 
Hadronic showers are regarded as jets if its deposited transverse energy $ E_{T} $ 
is greater than 10 GeV within a cone of radius $\Delta R=0.4$.
Electromagnetic calorimeters cells of dimensions $ \Delta \eta \times \Delta \phi
=0.025\times 0.025 $ within the same pseudorapidity range $ -2.5<\eta
<2.5 $ are used to measure the lepton energy. 
The electromagnetic energy resolution is given by $
0.1/\sqrt{E(GeV)}{\oplus }0.007 $  over this $ \eta $ region.
Electromagnetic showers are identified as electron candidates if
their $E_{T}>5 $~GeV within a cone of radius $ \Delta R=0.15$.
Default ATLFAST electron isolation criteria were applied: separation
by $\Delta R>0.4 $ from other clusters and  $E_{T}<10$~GeV 
deposition in  $\Delta R=0.2$ cone around the electron).     
Default selection criteria for muons ( \pT$>6$ GeV and $|\eta|< 2.5$ ) and 
isolation criteria, the same as for in the case of electron, were also applied. 

ATLFAST labels a jet as a $b$-jet if a $b$-quark is present in a cone $ \Delta
R=0.2 $ around the  reconstructed jet  for jets with $ \eta <2.5 $.
Efficiencies for b-jet tagging have been parametrized by
\pT-dependent  functions,  with maximal saturated efficiency 
$\epsilon_{b}=0.7$ at high \pT.
We have checked that for slepton mass within the range $100-1000$ GeV
and $\lambda^\prime > 0.005$, the lifetime of the neutralino
is shorter than the lifetime of a B-meson.
Therefore displacements of the B-meson vertex due
to the neutralino decay will not affect the b-tagging efficiency.

The analysis transverse momentum cuts used for electrons, muons and jets were
$10$ GeV, $10$ GeV and $20$ GeV, respectively.

\section{Signal and background analysis for $\tilde{\chi}^0\to cde$ channel}

\begin{figure}[htb] 
\vspace*{-0.5cm}
\epsfig{file=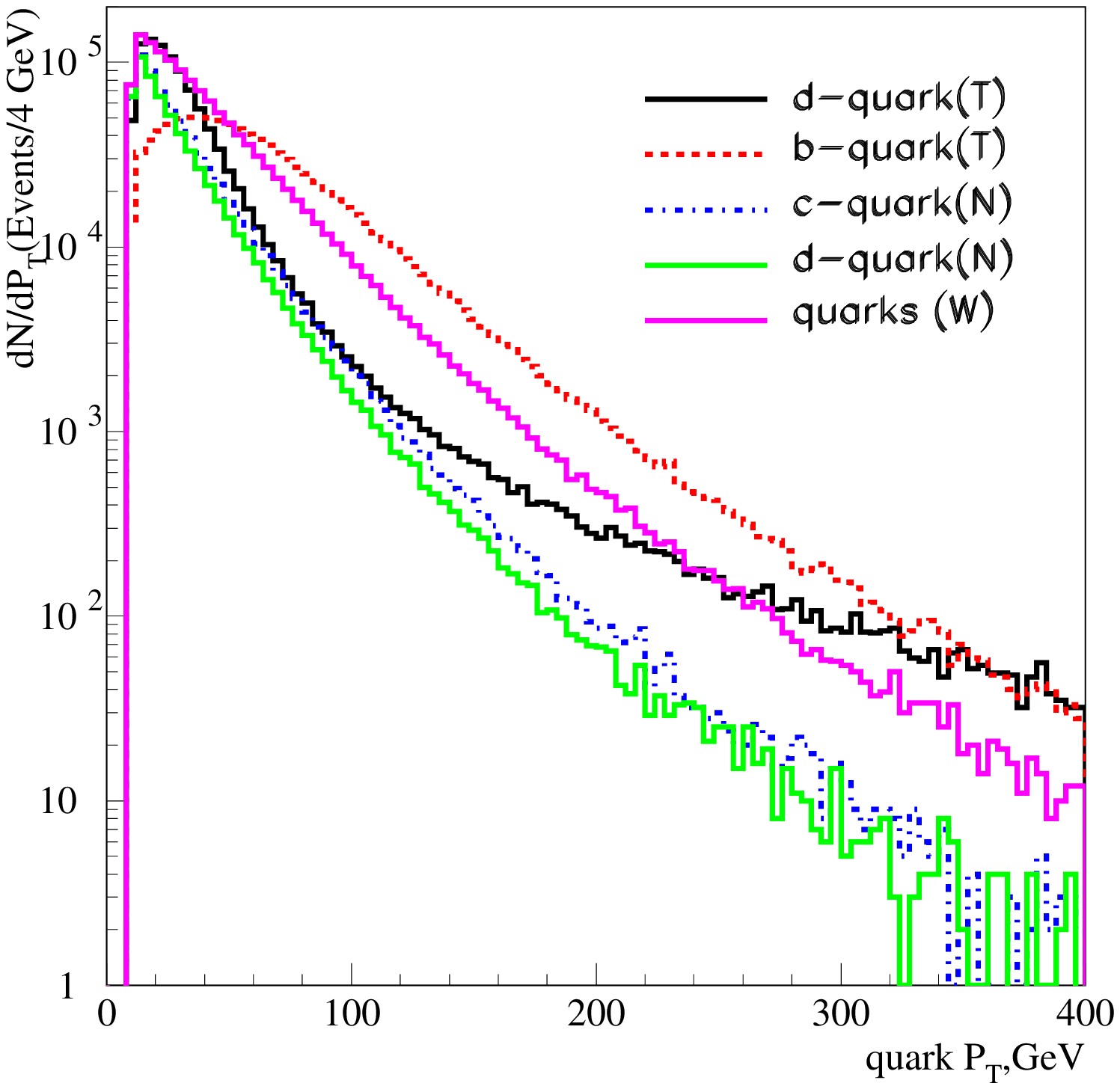,width=0.45\textwidth}
\epsfig{file=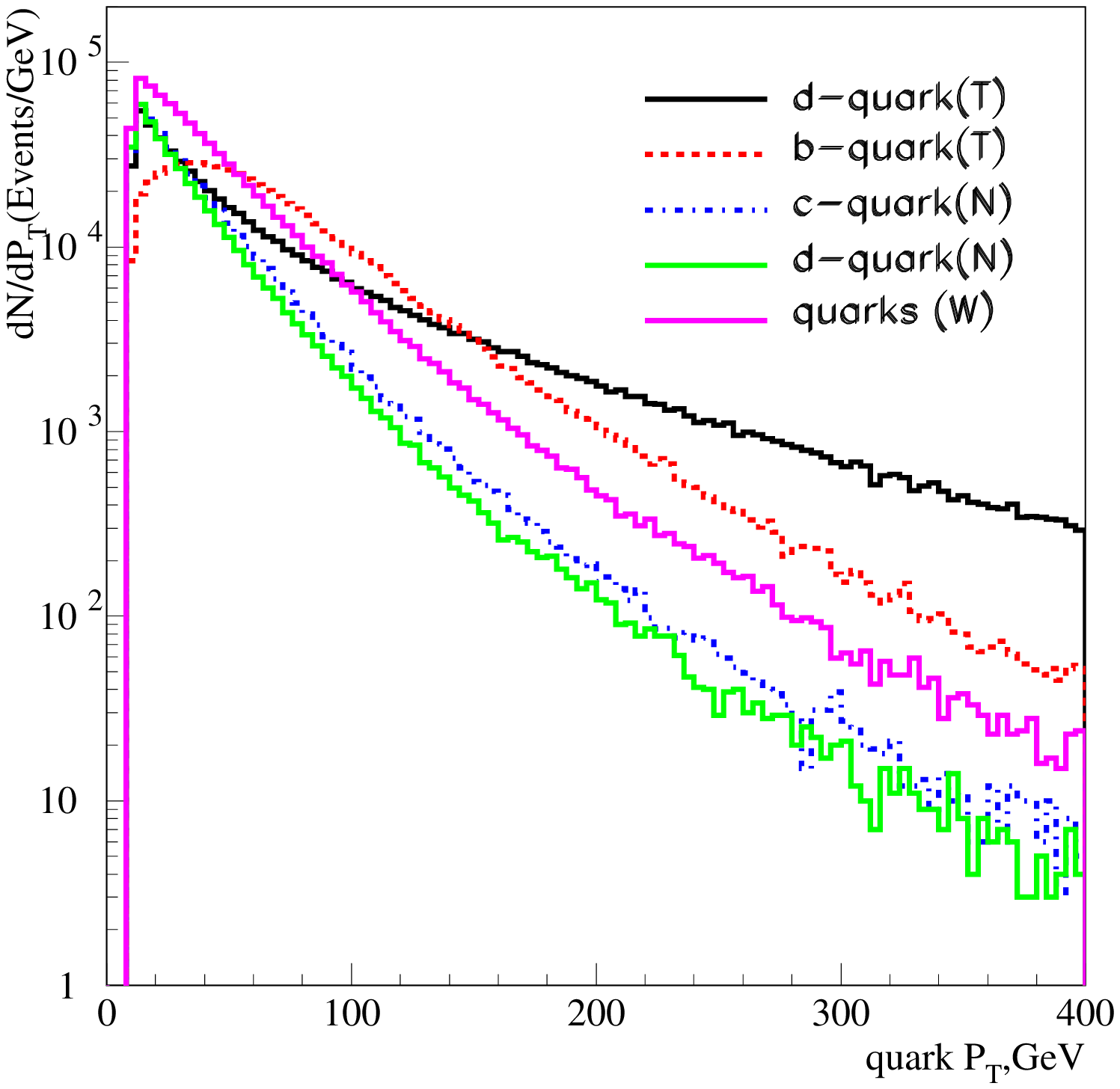,width=0.45\textwidth}
% %
\caption{Transverse momentum of quarks from the $pp \to t \tilde{\chi}^0 q\ell$ reaction
( $CDE$ channel ) for  $m_{\tilde\ell}=150$~GeV (left) and  $m_{\tilde\ell}=200$~GeV (right ).
(T) means a  quark originating from a top quark decay, whereas (N) denotes its
origin from a neutralino decay and (W) from the $W$ decay. 
\label{ptq}}
% %
\end{figure}

\begin{figure}[htb] 
\vspace*{-0.5cm}
\epsfig{file=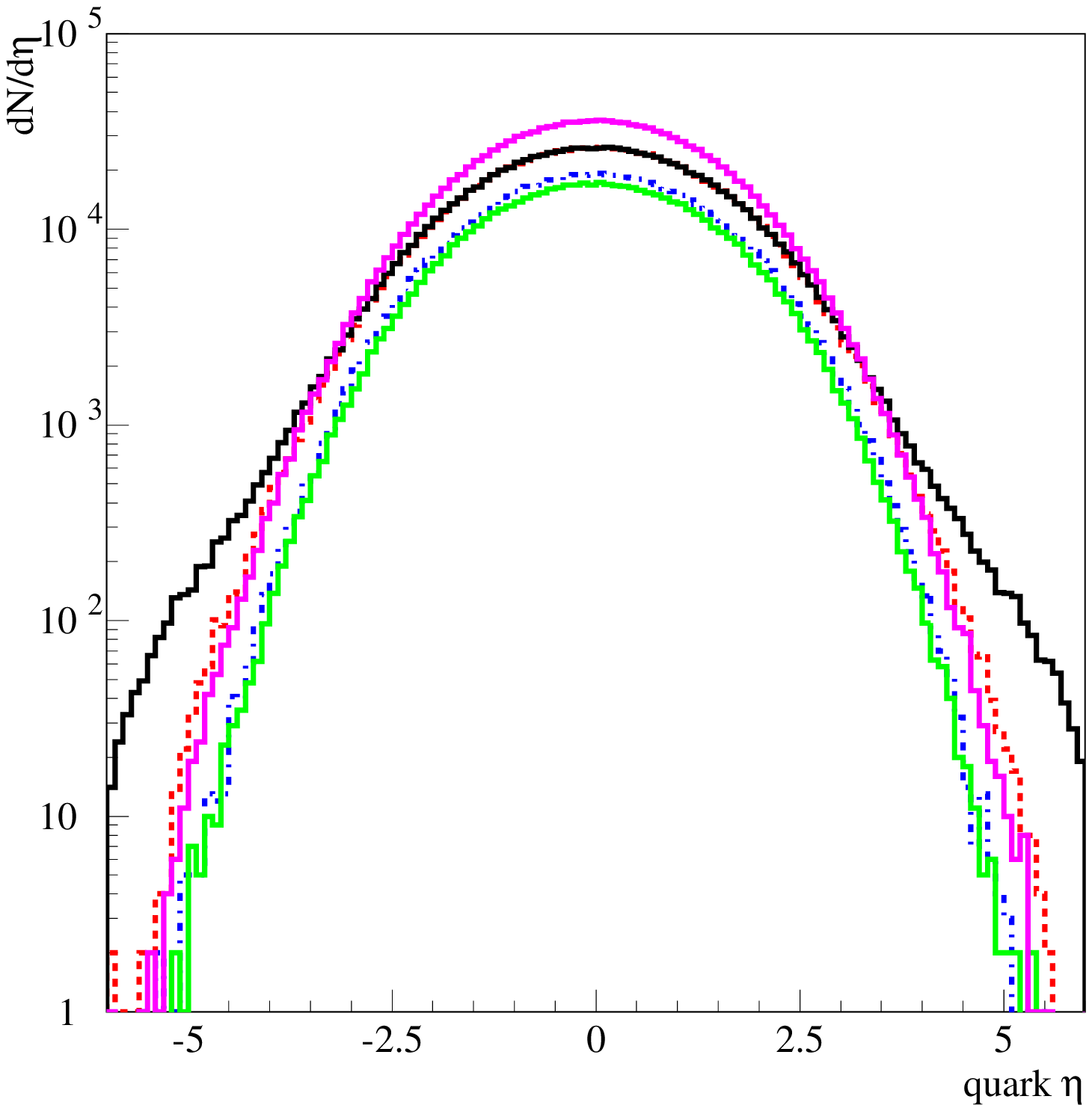,width=0.45\textwidth}
\epsfig{file=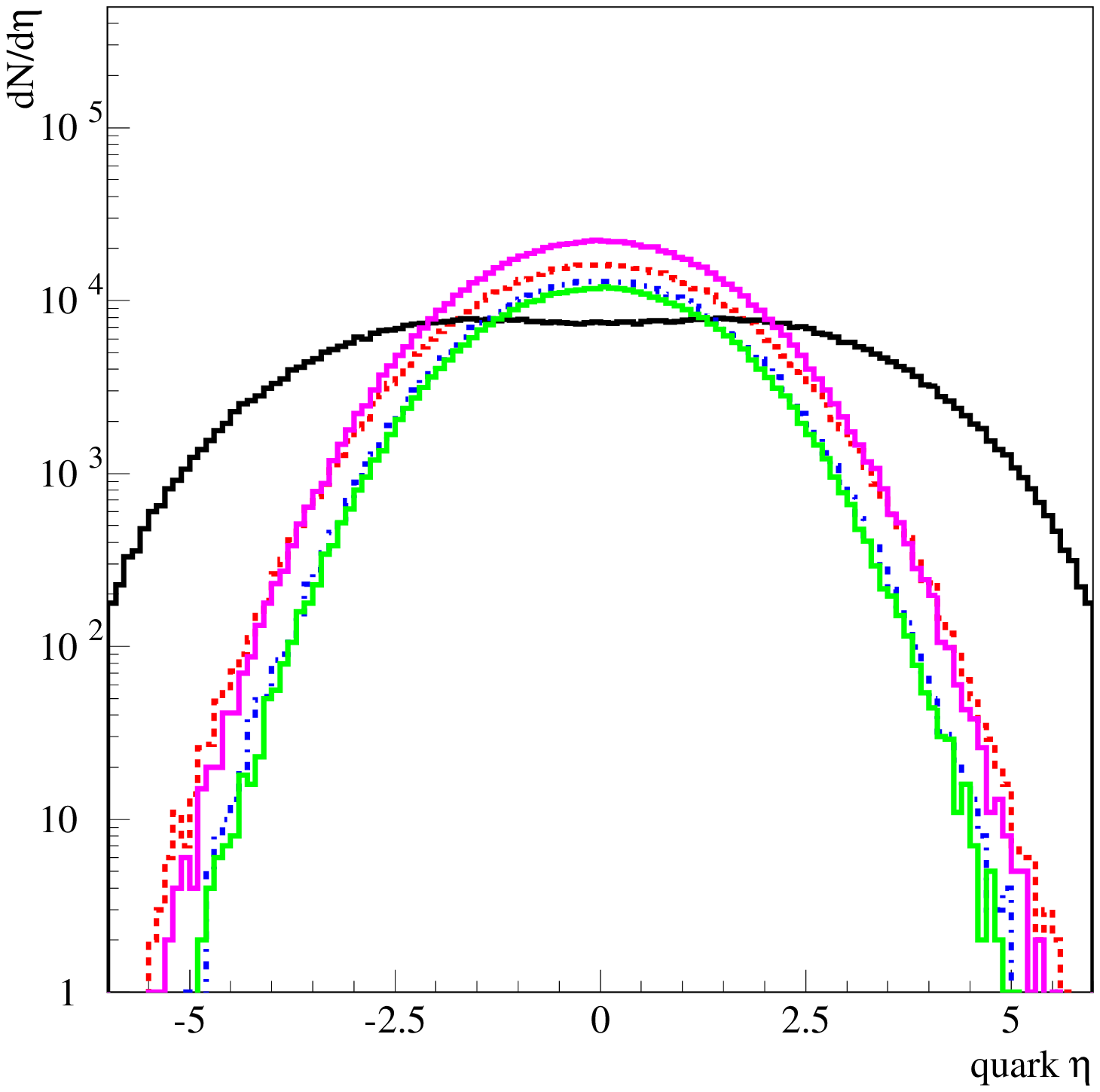,width=0.45\textwidth}
\caption{Pseudorapidity of quarks from the $pp \to t \tilde{\chi}^0 q\ell$ reaction for 
$m_{\tilde\ell}=150$~GeV (left) and  $m_{\tilde\ell}=200$~GeV (right). 
The notations are the same as in Fig.~\ref{ptq}.
\label{etq}}
\end{figure}

\begin{figure}[htb] 
\epsfig{file=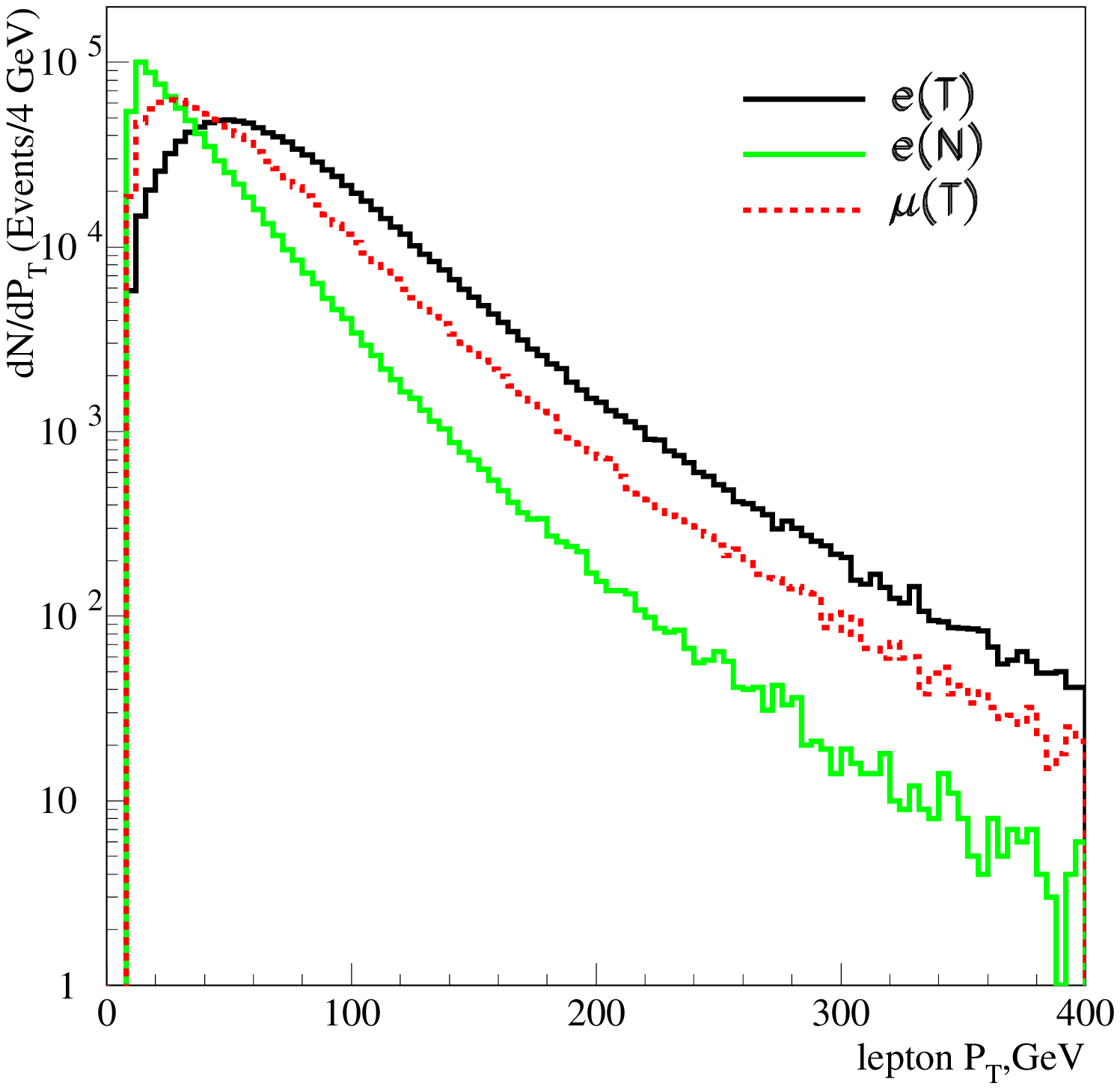,width=0.45\textwidth}
\epsfig{file=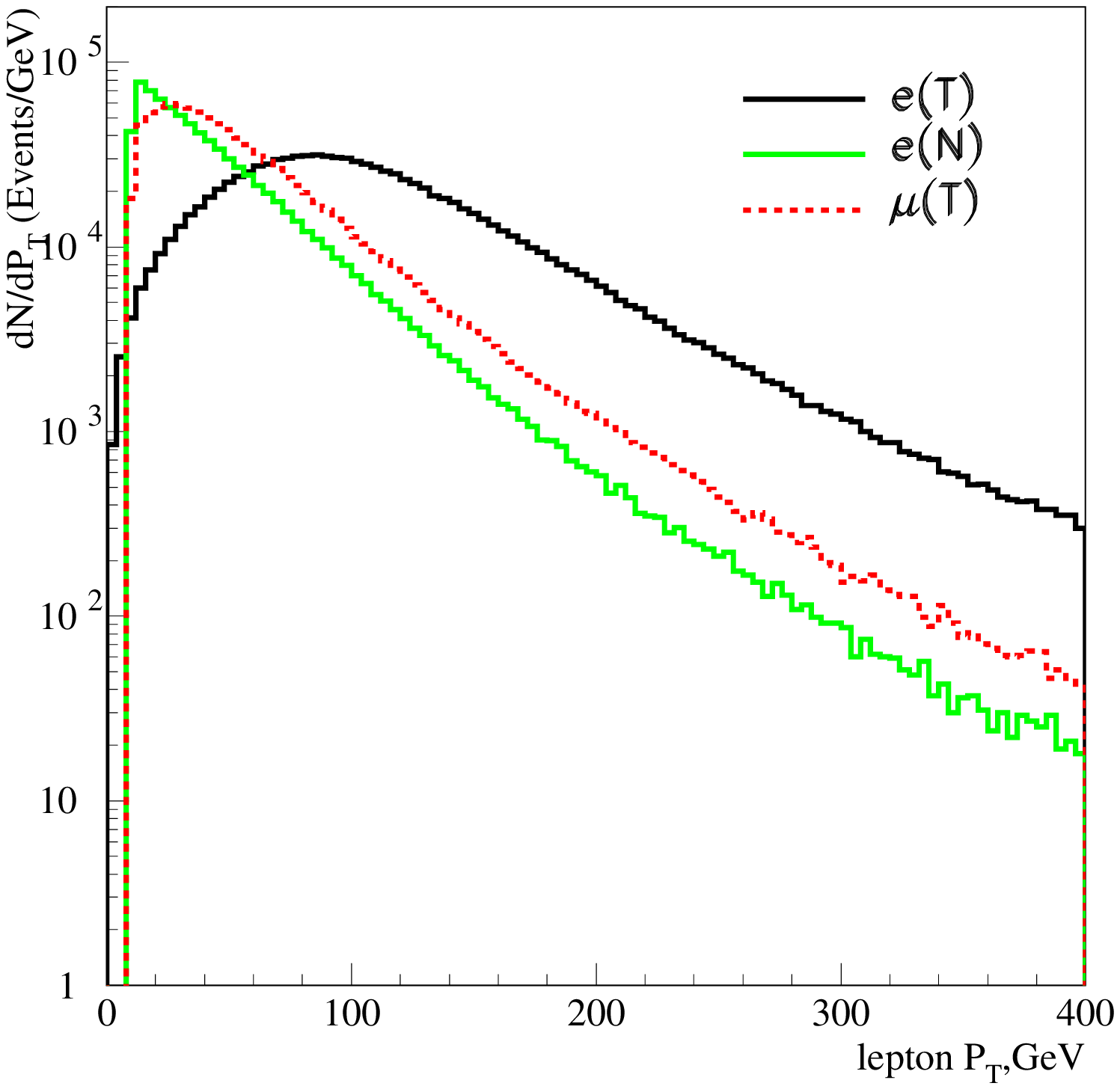,width=0.45\textwidth}
\caption{Transverse momentum of leptons from the $pp \to t \tilde{\chi}^0 q\ell$ reaction 
for $m_{\tilde\ell}=150$~GeV (left) and  $m_{\tilde\ell}=200$~GeV (right). 
(T) means a lepton originated from the top quark decay, whereas (N) denotes 
its origin from the neutralino decay and (W) from the $W$ decay.
\label{ptl}}
\end{figure}

\begin{figure}[htb] 
\epsfig{file=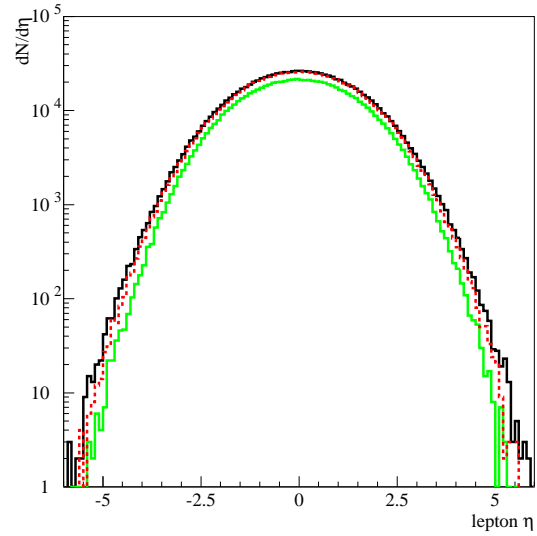,width=0.45\textwidth}%
\centering{
\caption{Pseudorapidity of leptons from the $pp \to t \tilde{\chi}^0 q\ell$ reaction for 
$m_{\tilde\ell}=150$~GeV.  
The notations are the same as in Fig.~\ref{ptl}.
\label{etl}}
}
\end{figure}

In this section we study $CDE$ neutralino decay channel: $\tilde{\chi}^0\to cde$.
First we analyze kinematical characteristics of jets for this channel.
The partonic distributions of the transverse momentum and pseudorapidity
of quarks and leptons for the process 
$pp \to t \tilde{\chi}^0 q\ell(\tilde{\chi}^0\to cde)$ are 
shown in Figs.~\ref{ptq},~\ref{etq},~\ref{ptl} and~\ref{etl}. 
For completeness, we also show distributions for jets
originating from $W \to jets$ decay from top-quark
undergoing SM decay mode.

%%%%%%%%%%%%%%%%%%%%%%%%%%%%%%%%%%%%%%%%%%%%%%%%%%%%%%%%%%%%%%%%%%%%  Fig.9
%%%%%%%%%%%%%%%%%%%%%%%%%%%%%%%%%%%%%%%%%%%%%%%%%%%%%%%%%%%%%%%%%%%%

The transverse momentum spectra of quarks in the process involving a light
slepton ($m_{\tilde\ell}=150$~GeV) (Fig.~\ref{ptq}(left)) indicate that the 
highest \pT jet in the  event,
with \pT $> 50$ GeV is a $b$-jet from the top quark SM decay.
Using $b$-jet tagging technique this jet can be effectively separated from the analysis.
Among jets of \pT $< 50$ GeV, a $d$-quark from the \rpv top quark decay dominates,
as well as in the very high \pT region.
For the case of a heavy slepton
($m_{\tilde\ell}=200$~GeV) , a $d$-quark from \rpv decay tends to be with the
highest \pT in the \pT $> 150$ GeV region and to be emitted in the forward
direction (see  Fig.~\ref{etq}(right)). The two jets with smallest \pT in the events, 
originate from a neutralino decay.  

%%%%%%%%%%%%%%%%%%%%%%%%%%%%%%%%%%%%%%%%%%%%%%%%%%%%%%%%%%%%%%%%%%%% Fig.10
%%%%%%%%%%%%%%%%%%%%%%%%%%%%%%%%%%%%%%%%%%%%%%%%%%%%%%%%%%%%%%%%%%%%

The transverse momentum for a lepton from \rpv top quark decay is the largest
in the region \pT $> 50$ GeV and \pT $> 75$ GeV for the light and heavy 
slepton set of events, respectively (see Fig.~\ref{ptl}).
In terms of the pseudorapidity variable, leptons are mainly emitted in the central region
(Fig.~\ref{etl}).

%%%%%%%%%%%%%%%%%%%%%%%%%%%%%%%%%%%%%%%%%%%%%%%%%%%%%%%%%%%%%%%%%%%% Fig.11
%%%%%%%%%%%%%%%%%%%%%%%%%%%%%%%%%%%%%%%%%%%%%%%%%%%%%%%%%%%%%%%%%%%%

%%%%%%%%%%%%%%%%%%%%%%%%%%%%%%%%%%%%%%%%%%%%%%%%%%%%%%%%%%%%%%%%%%%% Fig.12

%%%%%%%%%%%%%%%%%%%%%%%%%%%%%%%%%%%%%%%%%%%%%%%%%%%%%%%%%%%%%%%%%%%%%

\begin{figure}[htb] 
\epsfig{file=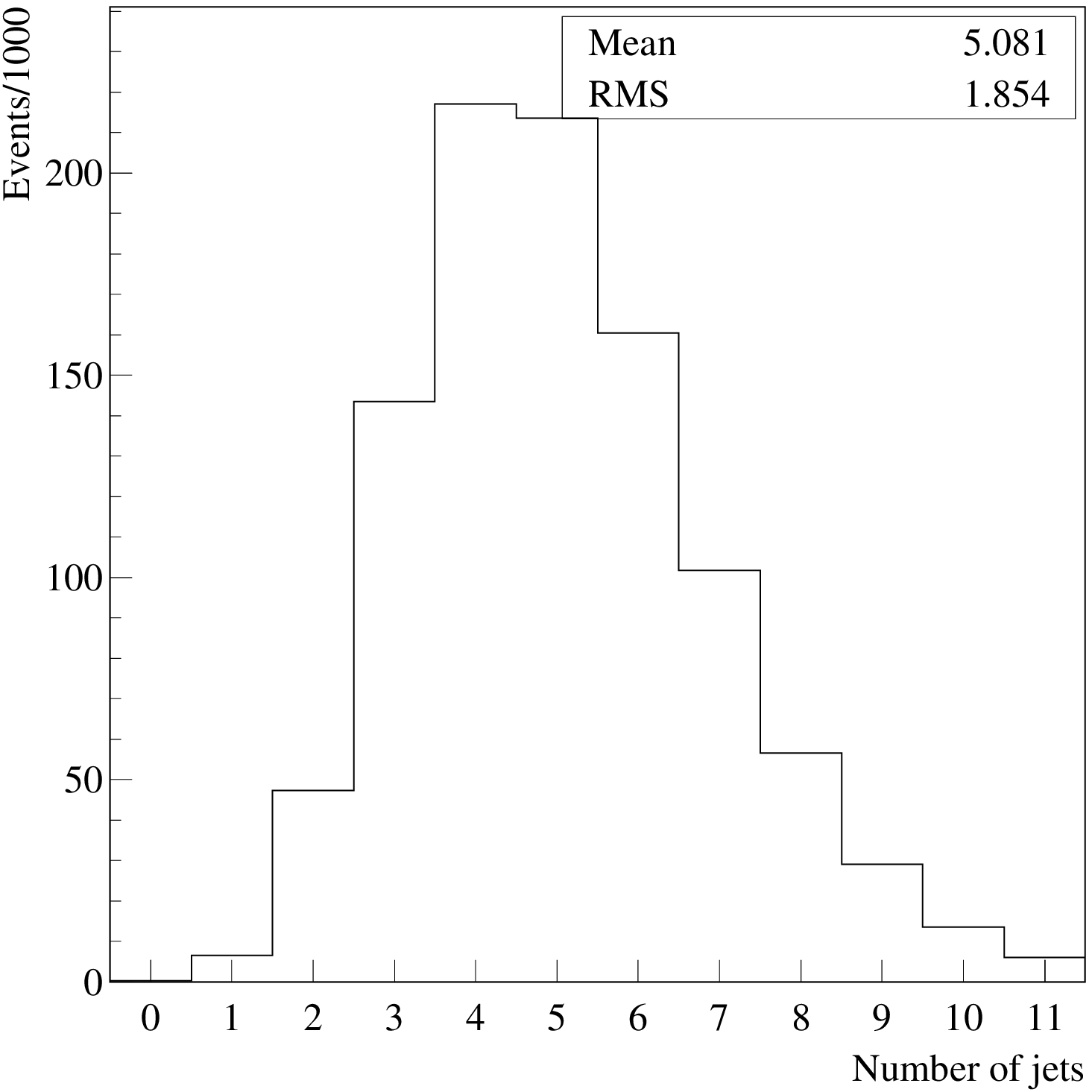,width=0.45\textwidth}
\epsfig{file=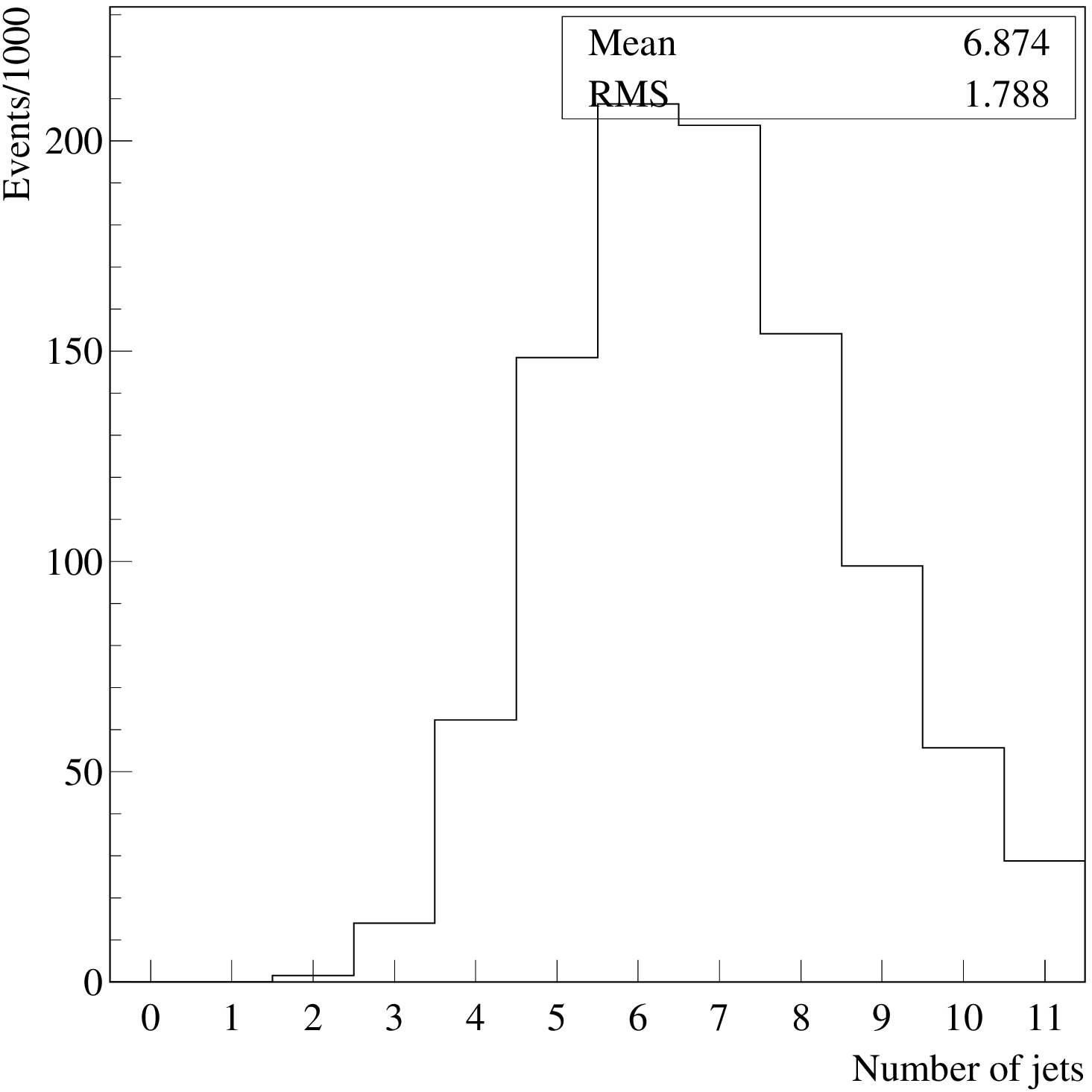,width=0.45\textwidth}
\caption{Distribution of the number of reconstructed jets after
ATLFAST for the cases of $W\to \mu\nu$ (left) and $W\to jets$. 
\label{njet}}
\end{figure}

The effect of initial (ISR) and final state (FSR) QCD radiation 
approximately doubles the number of jets with \pT$>10$ GeV
produced in the event. Fig.~\ref{njet} shows the jet multiplicity distribution.

%%%%%%%%%%%%%%%%%%%%%%%%%%%%%%%%%%%%%%%%%%%%%%%%%%%%%%%%%%%%%%%%%%%%   Fig.13
%%%%%%%%%%%%%%%%%%%%%%%%%%%%%%%%%%%%%%%%%%%%%%%%%%%%%%%%%%%%%%%%%%%%

\begin{figure}[htb] 
\epsfig{file=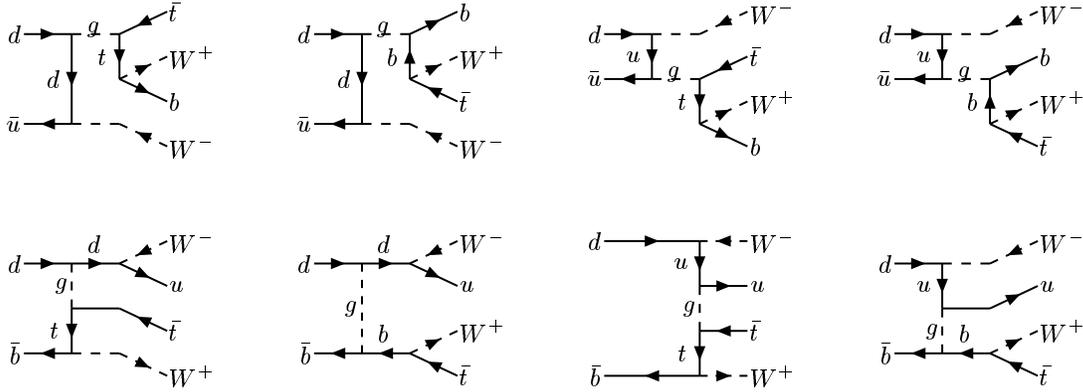,width=\textwidth}
\caption{Tree-level diagrams for the $tWWq$ process
\label{twwj}}
\end{figure}

The only non-negligible background for the $CDE$ channel is $tWW(q)$ processes.
This process would be  a background
for the case where $\tilde{\chi}^0\to cde$ when
both $W$'s decay leptonically and the $W$ from the top quark decays
hadronically. 
The Feynman diagrams for $tWWq$ background
are shown in  Fig.~\ref{twwj}.
The total cross-section for this process, obtained from CompHEP is 1.1 pb. 
 
%%%%%%%%%%%%%%%%%%%%%%%%%%%%%%%%%%%%%%%%%%%%%%%%%%%%%%%%%%%%%%%%%%%% Fig.14
%%%%%%%%%%%%%%%%%%%%%%%%%%%%%%%%%%%%%%%%%%%%%%%%%%%%%%%%%%%%%%%%%%%%

\section{Results for $\tilde{\chi}^0\to cde$ channel}

The following cuts have 
been worked out for the top quark reconstruction:

\begin{itemize}

\item At least 4 ( or 6 for $t\to bjj$ final state ) jets with \pT $> 20$ GeV 
      and one electron with \pT $> 10$ GeV within $|\eta| < 2.5$ pseudorapidity.

\item one jet tagged as a $b$-jet: required to get rid of a $b$-jet from a top quark
      which decayed in the framework of the SM.

\item cut on invariant mass of ($jjee$), $M(jjee) < 150$ GeV (for the light 
      slepton case) or $< 200$ GeV (for the heavy slepton case),
      to suppress the influence of the on-shell slepton on the top quark signal, 
      especially effective in the $m_{\tilde\ell}< m_{top}$ case.

\end{itemize}

      For the mass reconstruction we combine two electrons with the three 
      jets in the event. 
      If the number of jets in the event was larger than 
      three (but less than seven), the three jets with the smallest \pT in the event 
      were combined with both electrons.

\begin{figure}[htb] 
\epsfig{file=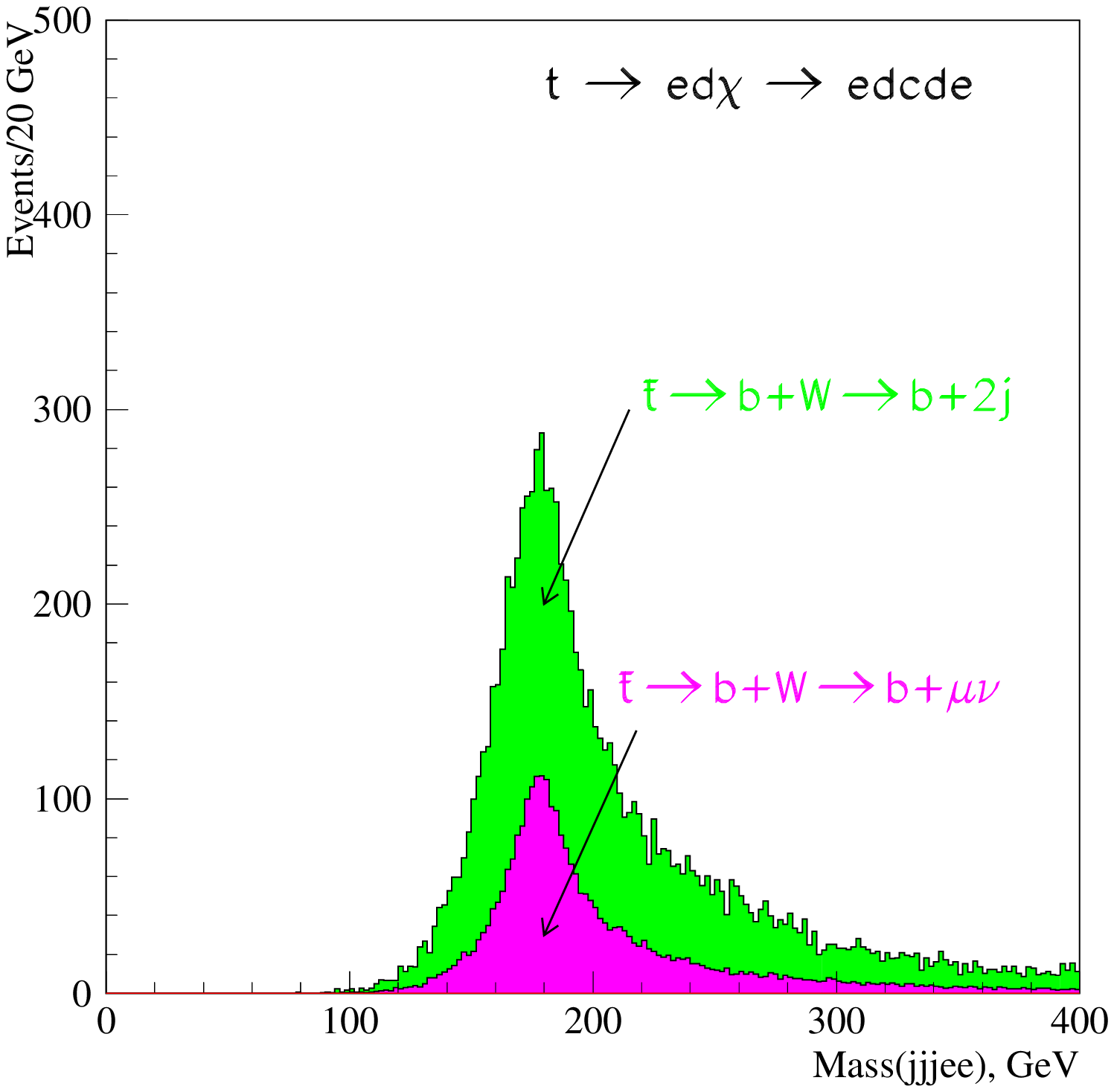,width=0.45\textwidth}
\epsfig{file=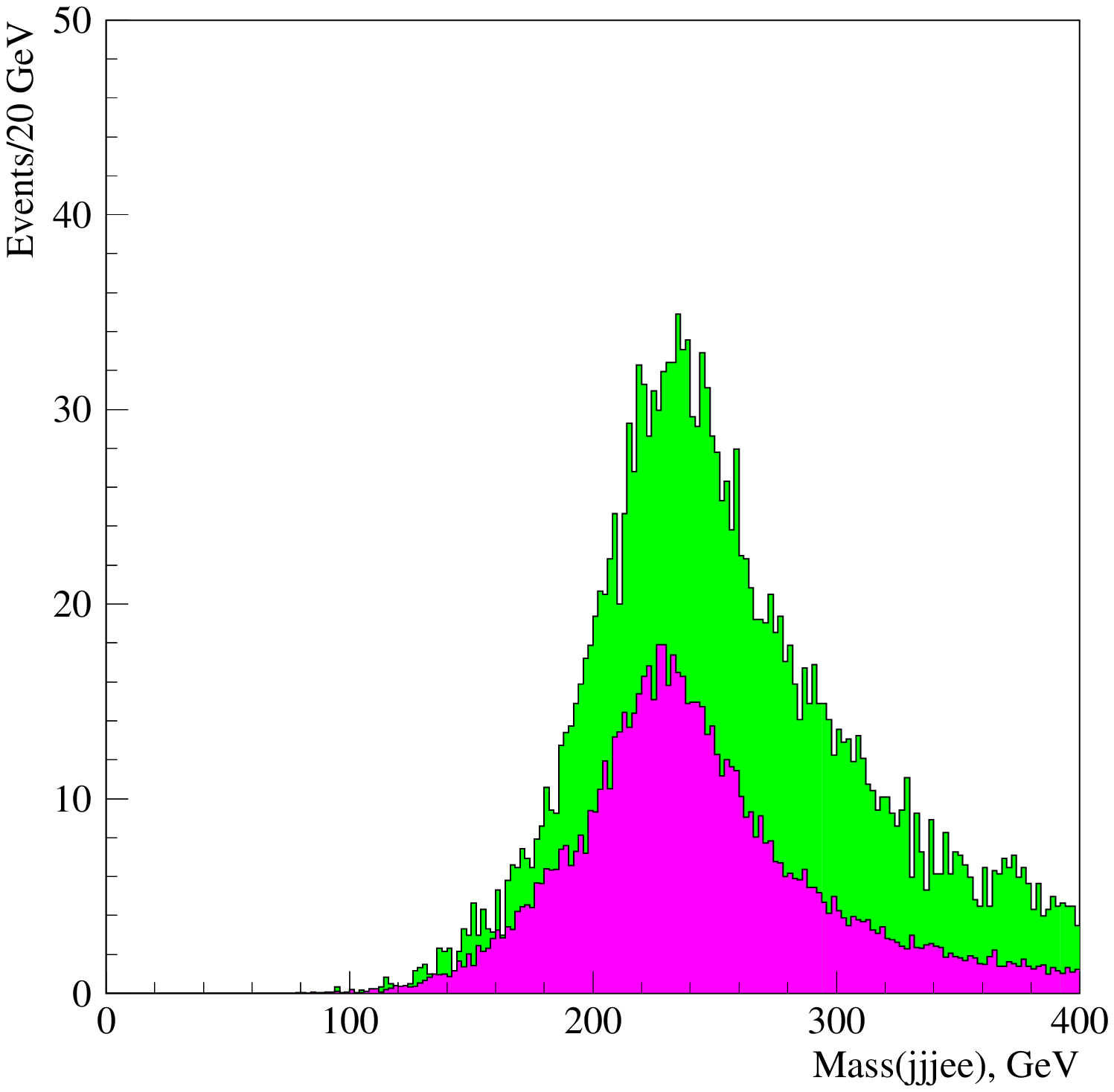,width=0.45\textwidth}
\caption{Detector level invariant mass distribution for the (jjjee) system in $CDE$ channel and light (left) or heavy (right) slepton cases (see text). The distributions are normalized for an integrated luminosity of $100 fb^{-1}$ and $300 fb^{-1}$, respectively.
\label{masstop1}}
\end{figure}

The resulting mass distributions of the ($jjjee$) system are shown in Fig.~\ref{masstop1}
for the two different $W$ decay modes studied and the light slepton case , $m_{\tilde\ell}=150$~GeV 
(left), and the heavy one, $m_{\tilde\ell}=200$~GeV (right), respectively.
The data are presented for
an integrated luminosity of 100 fb$^{-1}$ (one year of running at high luminosity at LHC)
for the light slepton case, and 300 fb$^{-1}$ for the heavy slepton case.
After the fiducial cuts given above, the $tWWq$  background for the events with
$\tilde{\chi}^0\to cde$ is appears to be very small and practically is not seen on the 
Fig.~\ref{masstop1}.

The events for the heavy slepton case show the shift of the mass value of
the top quark signature because of a heavy on-shell slepton mass dominance and
big combinatorial jet pairs background.

%%%%%%%%%%%%%%%%%%%%%%%%%%%%%%%%%%%%%%%%%%%%%%%%%%%%%%%%%%%%%%%%%%%%  Fig.15
%%%%%%%%%%%%%%%%%%%%%%%%%%%%%%%%%%%%%%%%%%%%%%%%%%%%%%%%%%%%%%%%%%%%

Based on the kinematical distributions shown above, we define
procedures to select and reconstruct the 
neutralino and the slepton in the events of interest.

\begin{figure}[htb] 
\epsfig{file=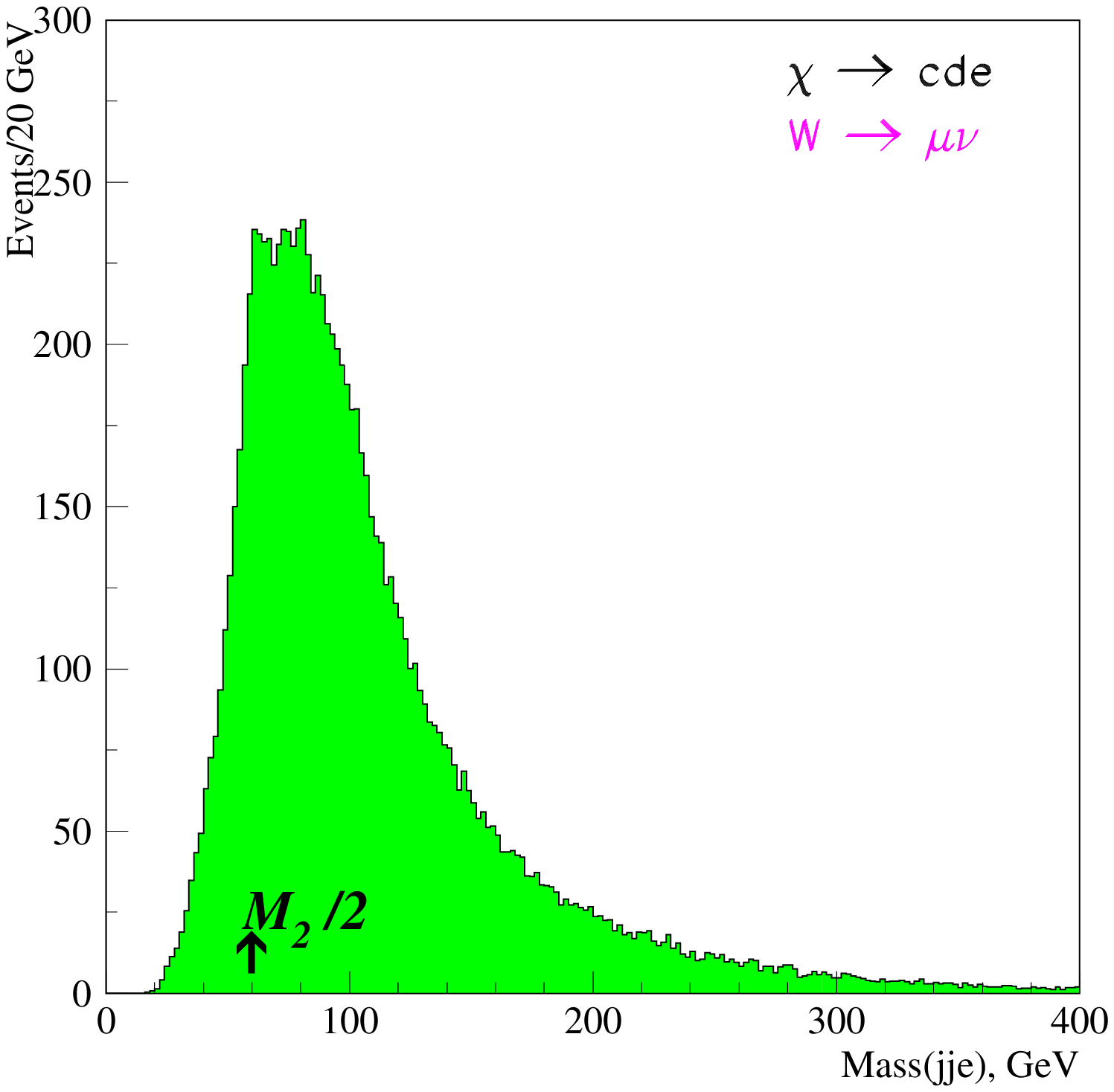,width=0.45\textwidth}
\epsfig{file=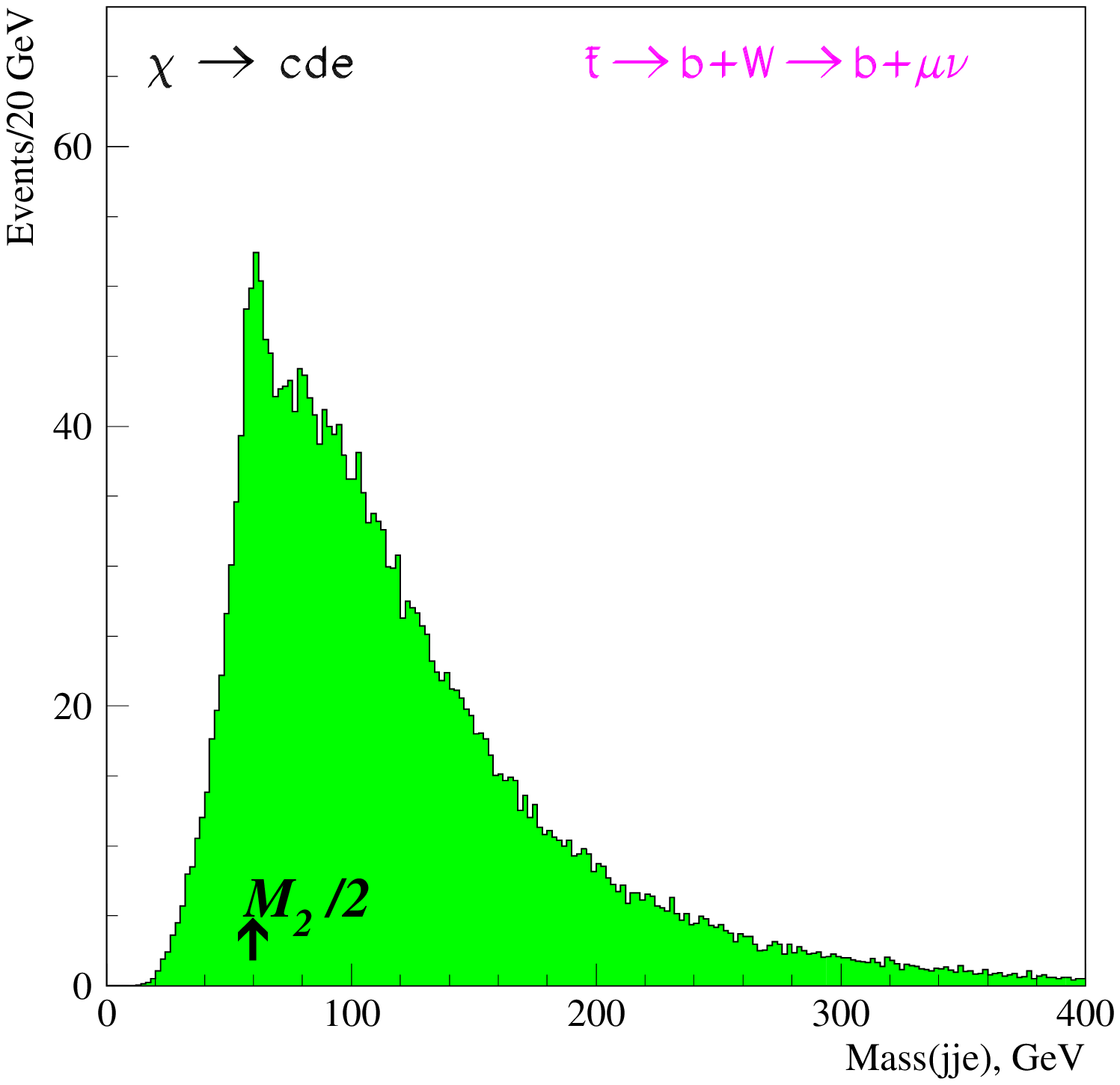,width=0.45\textwidth}
\caption{Detector level invariant mass distribution for the (jje) system for $CDE$ channel and the light slepton (left) or the heavy slepton ( right) cases. The distributions are normalized for an integrated luminosity of $100 fb^{-1}$ and $300 fb^{-1}$, respectively.
\label{massns1}}
\end{figure}

For the case of a light slepton,
in order to reconstruct the neutralino  (Fig.~\ref{massns1}, left), 
we combine the electron with smallest \pT with the two jets with smallest \pT.
The figure shows the invariant mass distribution for an integrated 
luminosity of
$100 fb^{-1}$, with the contributions from events where $W \to \mu\nu$ 
and where $W \to jets$ explicitly indicated. 

For the heavy slepton case (Fig.~\ref{massns1}, right) we use the same
procedure, but due to a lower production cross-section, we
present the mass reconstruction for  an integrated 
luminosity of $300 fb^{-1}$.

%%%%%%%%%%%%%%%%%%%%%%%%%%%%%%%%%%%%%%%%%%%%%%%%%%%%%%%%%%%%%%%%%%%% Fig.16
%%%%%%%%%%%%%%%%%%%%%%%%%%%%%%%%%%%%%%%%%%%%%%%%%%%%%%%%%%%%%%%%%%%%

\begin{figure}[htb] 
\epsfig{file=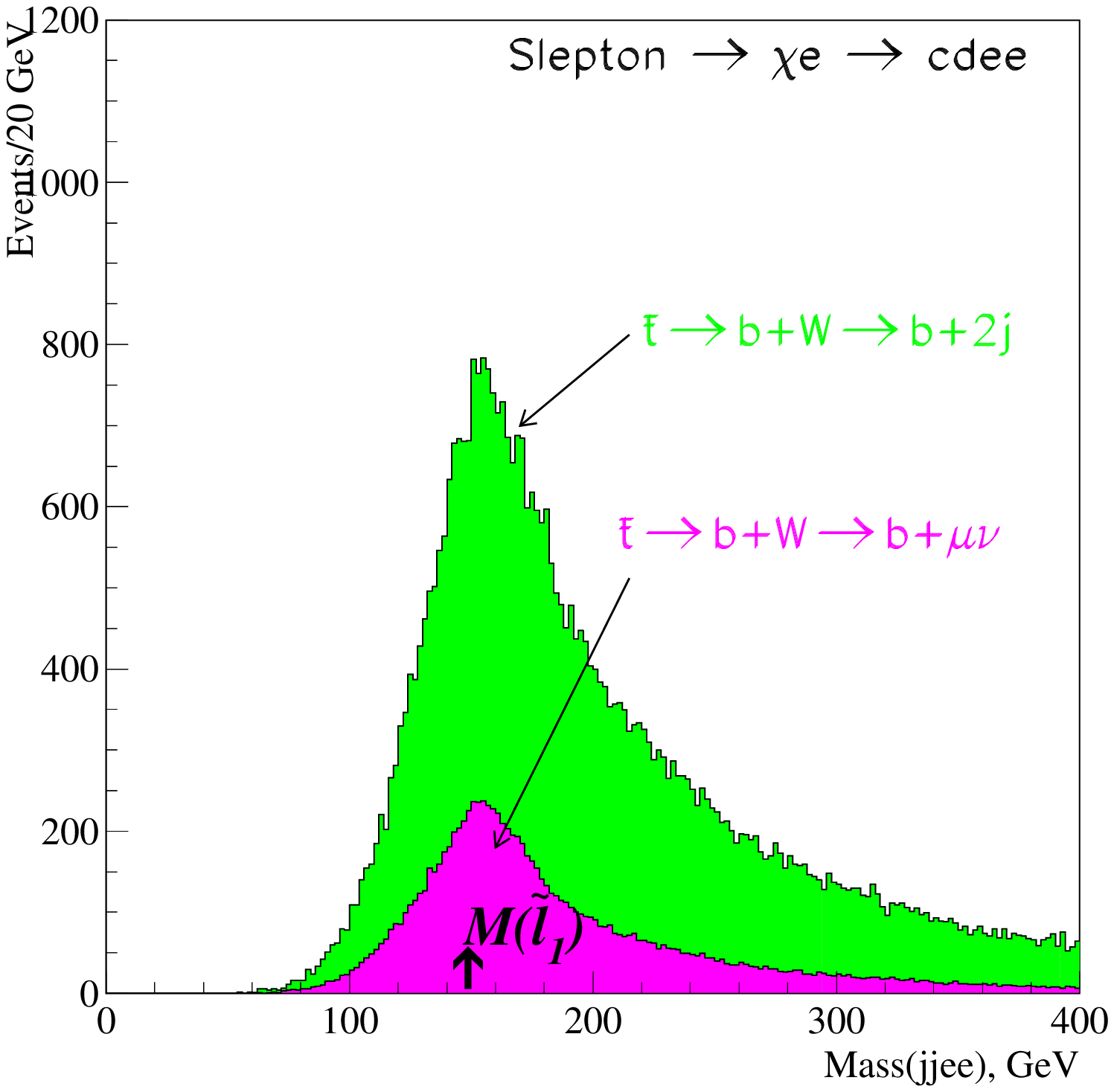,width=0.45\textwidth}
\epsfig{file=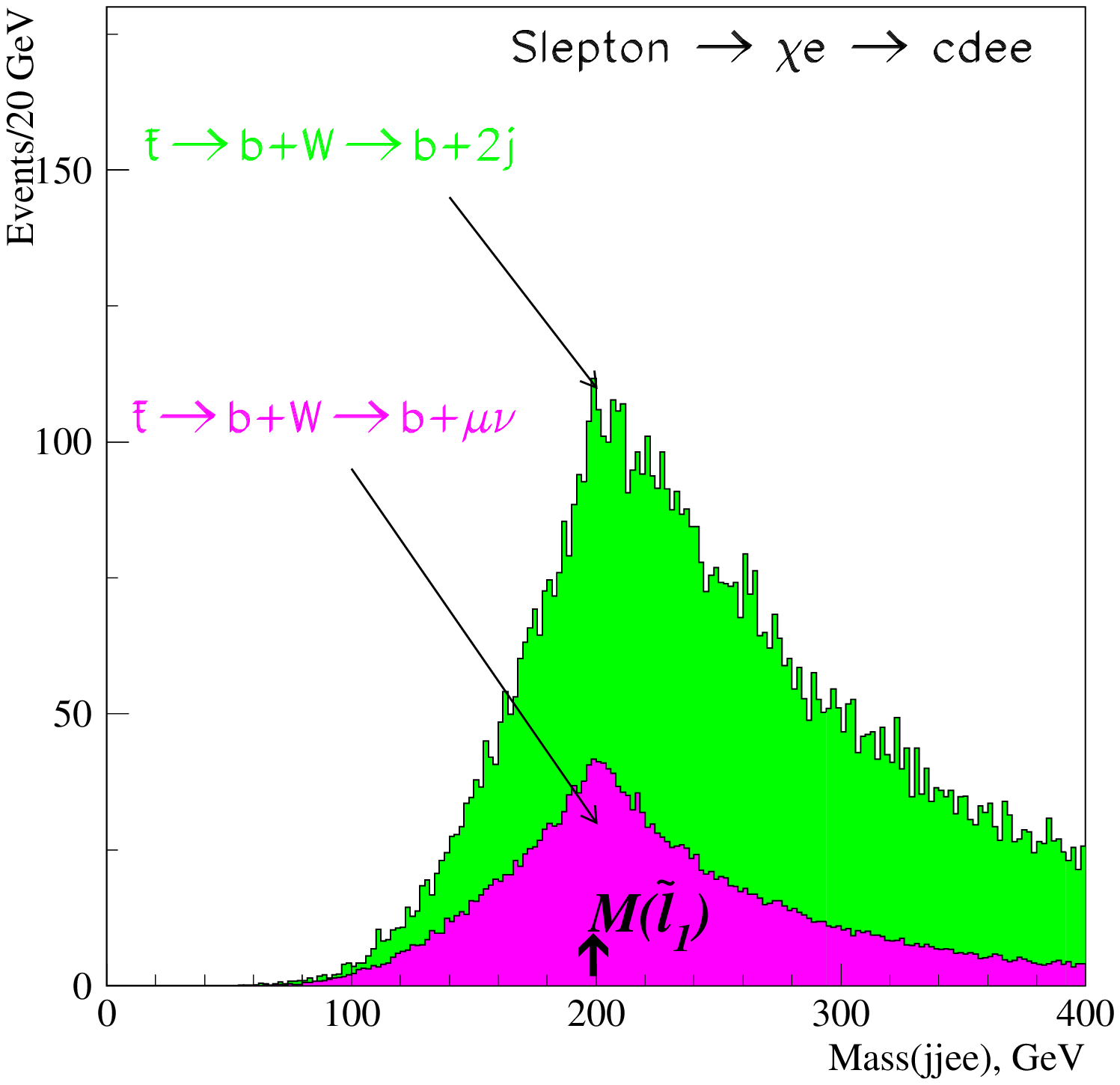,width=0.45\textwidth}
\caption{Detector level invariant mass distribution for the (jjee) system for $\tilde{\chi}^0 \rightarrow cde$ and the light slepton (left) or the heavy slepton ( right) cases. The distributions are normalized for an integrated luminosity of $100 fb^{-1}$ and $300 fb^{-1}$, respectively.
\label{massns2}}
\end{figure}

For the reconstruction of the slepton, we combine both electrons in the 
event with the same two soft jets (Fig.~\ref{massns2}).

%%%%%%%%%%%%%%%%%%%%%%%%%%%%%%%%%%%%%%%%%%%%%%%%%%%%%%%%%%%%%%%%%%%% Fig.17
%%%%%%%%%%%%%%%%%%%%%%%%%%%%%%%%%%%%%%%%%%%%%%%%%%%%%%%%%%%%%%%%%%%%

Light slepton case presented in Fig.~\ref{massns2} (left). 

For the heavy slepton case (Fig.~\ref{massns2}, right) we use the same
procedure, but again due to a lower production cross-section, we
present the mass reconstruction of sleptons for an integrated 
luminosity of $300 fb^{-1}$.  

%%%%%%%%%%%%%%%%%%%%%%%%%%%%%%%%%%%%%%%%%%%%%%%%%%%%%%%%%%%%%%%%%%%%%%%%%%%%%%%% BDN starts here

\section{Results for $\tilde{\chi}^0\to bd\nu$ channel}

\begin{figure}[htb] 
\epsfig{file=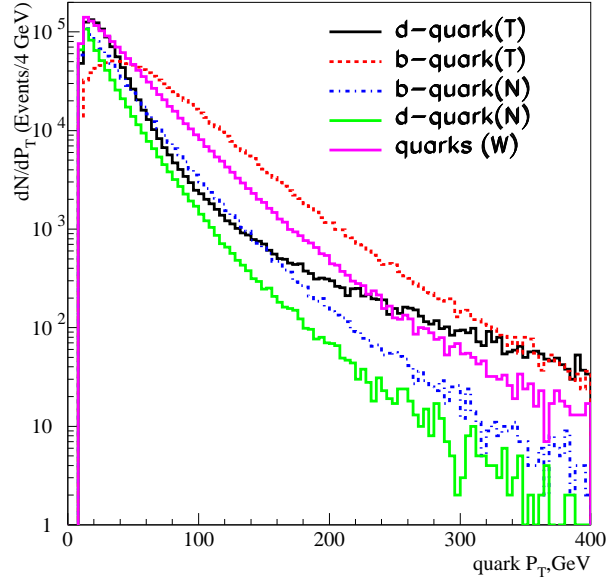,width=0.5\textwidth}
\centering{
\caption{Transverse momentum 
of quarks from the $pp \to t \tilde{\chi}^0 q\ell$
reaction for  $m_{\tilde\ell}=150$~GeV where $\tilde{\chi}^0 \rightarrow  bd\nu$. 
(T) means a quark originated from a top quark decay, whereas (N)  denotes its
origin from a neutralino decay and (W) from a $W$ decay. 
\label{ptq1}}
}
\end{figure}

In the case where $\tilde{\chi}^0\to bd\nu$, we considered only the scenario
where $W$ decays into jets, since the decay mode $W\to\nu\ell$ 
leads to two  neutrinos in the final state, making the event reconstruction
difficult. Fig.~\ref{ptq1} shows the partonic distributions of the transverse 
momentum of quarks for events where the neutralino decays 
$\tilde{\chi}^0\to bd\nu$(right). 

%%%%%%%%%%%%%%%%%%%%%%%%%%%%%%%%%%%%%%%%%%%%%%%%%%%%%%%%%%%%%%%%%%%%  Fig.18
%%%%%%%%%%%%%%%%%%%%%%%%%%%%%%%%%%%%%%%%%%%%%%%%%%%%%%%%%%%%%%%%%%%%

Thus, we are led to event topologies  with 6 quarks, one electron and a neutrino 
at the parton level.
Although the large combinatorial background in jet pair selection 
makes it difficult to reconstruct the signal for the event under study, 
one can observe certain correlations in these distributions. 
For example, the two softest light jets are originating from a
neutralino. 

The natural background for the case where $\tilde{\chi}^0\to bd\nu$ is \ttb and, 
possibly, \ttbb production.
These backgrounds have the topology of the signal when one $W$-boson decays into jets and
the other $W$ decays leptonically to $e\nu$.

We applied the following selection cuts for reconstructing events where one top quark 
decays via the \rpv SUSY mechanism, and its decay product, a neutralino, decays to $bd\nu$, 
while the other top quark in the event, decays in the SM to a $b$ quark and two jets.

\begin{center}
\begin{figure}[htb] 
\epsfig{file=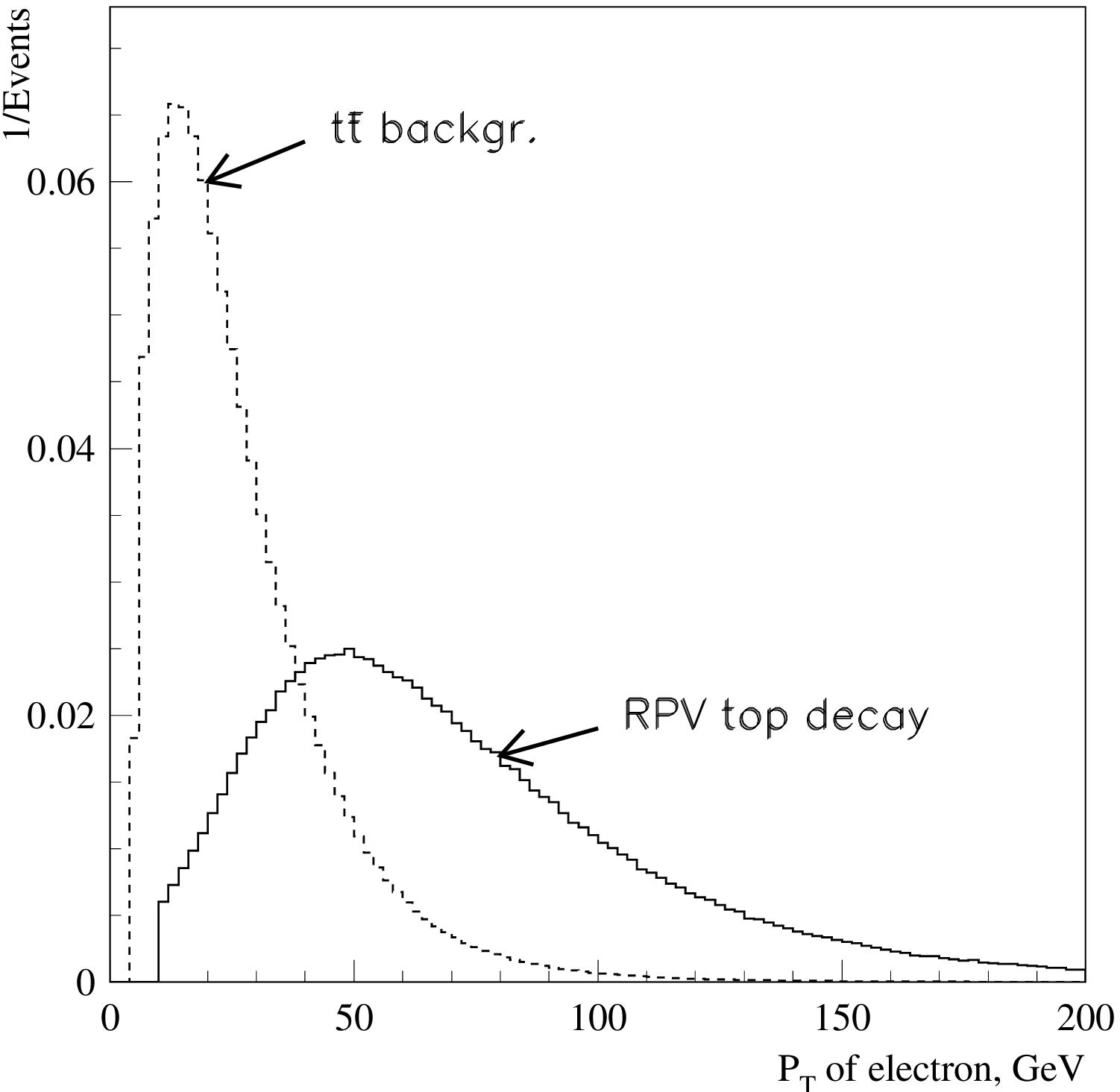,width=0.5\textwidth}%
\epsfig{file=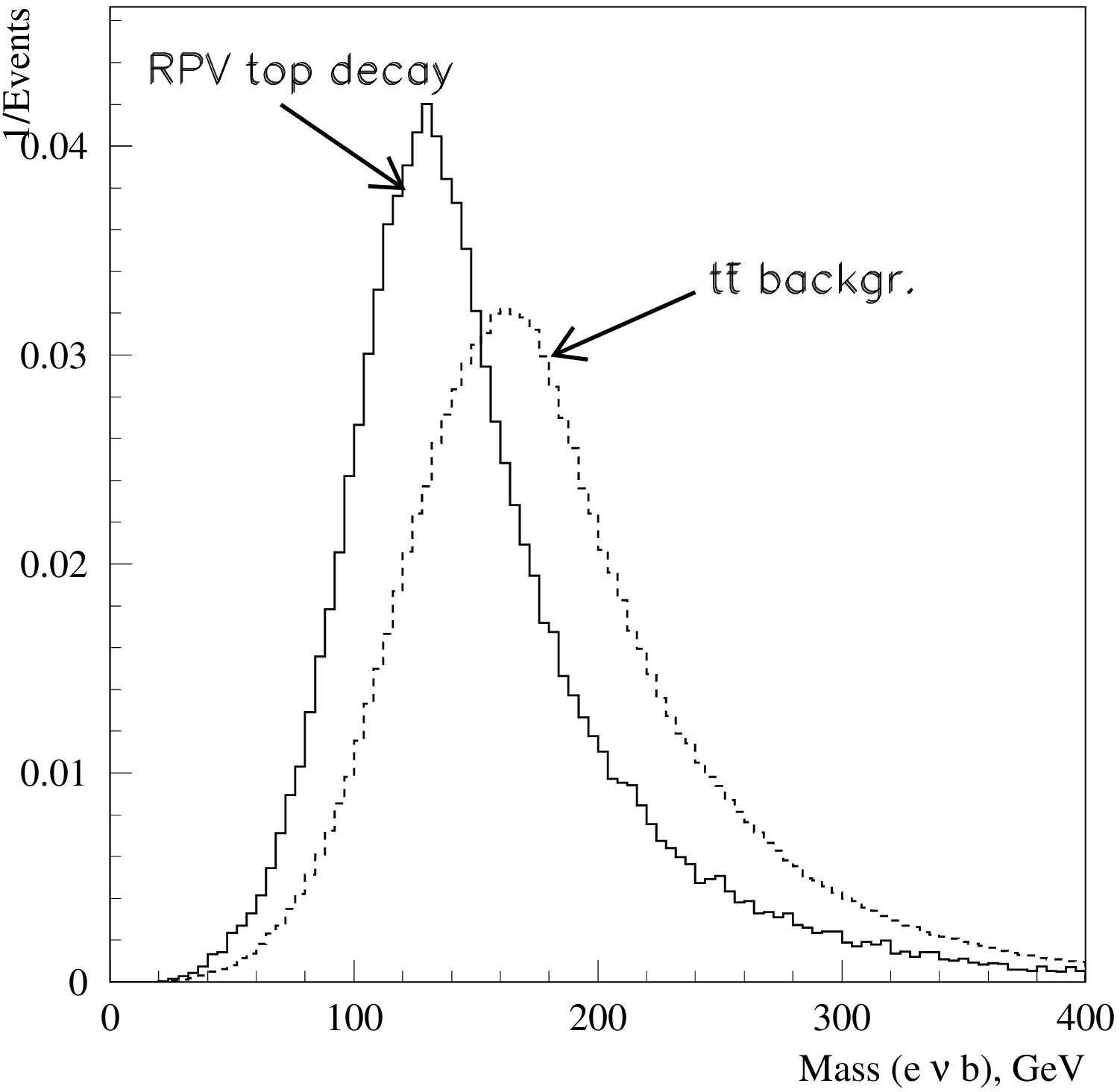,width=0.5\textwidth}
\caption{Parton level transverse momentum of electron produced in \rpv top decay (solid line)
and in $W$-boson leptonic decay in \ttb events (dashed line) (left); 
Detector level invariant mass distribution for 
($e \nu b$-jet) system in the \rpv top decay events with a neutralino decay to $bd\nu$
(solid line) and \ttb events (dashed line )(right).  
\label{wtm}}
\end{figure}
\end{center}

\begin{itemize}

\item at least 6 jets with \pT $> 20$ GeV and one electron
         with \pT $> 40$ GeV and $|\eta| < 2.5$ (see Fig.\ref{wtm}).

\item at least two jets tagged as b-jets with \pT $> 20$ GeV:
       one for the top quark with the SM decay and another
       from the decay of the neutralino. 

\item No two-jet combination should have the invariant mass of a $W$, in a window 
      $\pm 20$ GeV.

\item cut on the mass of the system ($e b \nu)< 140$ GeV (see Fig.\ref{wtm}), 
      to suppress the \ttb background, since no such decays of the top quark
      should occur in the \rpv top decay scenario.

\item cut on the transverse mass of ($jje\nu$): $M(jje\not\hskip-5truedd E_{T}) < 150$ GeV
      (for the light slepton case )  to suppress the influence of the on-shell slepton 
      on the top quark signal, especially effective in the $m_{\tilde\ell}< m_{top}$ 
      case.

\end{itemize}

      The mass reconstruction was performed for one electron, \emiss  and three 
      jets in the event. When the number of jets in the event was larger than 
      three (but less than seven), the three jets with the smallest \pT in the event 
      were taken into the reconstruction system. 

%%%%%%%%%%%%%%%%%%%%%%%%%%%%%%%%%%%%%%%%%%%%%%%%%%%%%%%%%%%%%%%%%%%% Fig.19
%%%%%%%%%%%%%%%%%%%%%%%%%%%%%%%%%%%%%%%%%%%%%%%%%%%%%%%%%%%%%%%%%%%%

\begin{figure}[htb]
\vspace*{-0.5cm} 
\epsfig{file=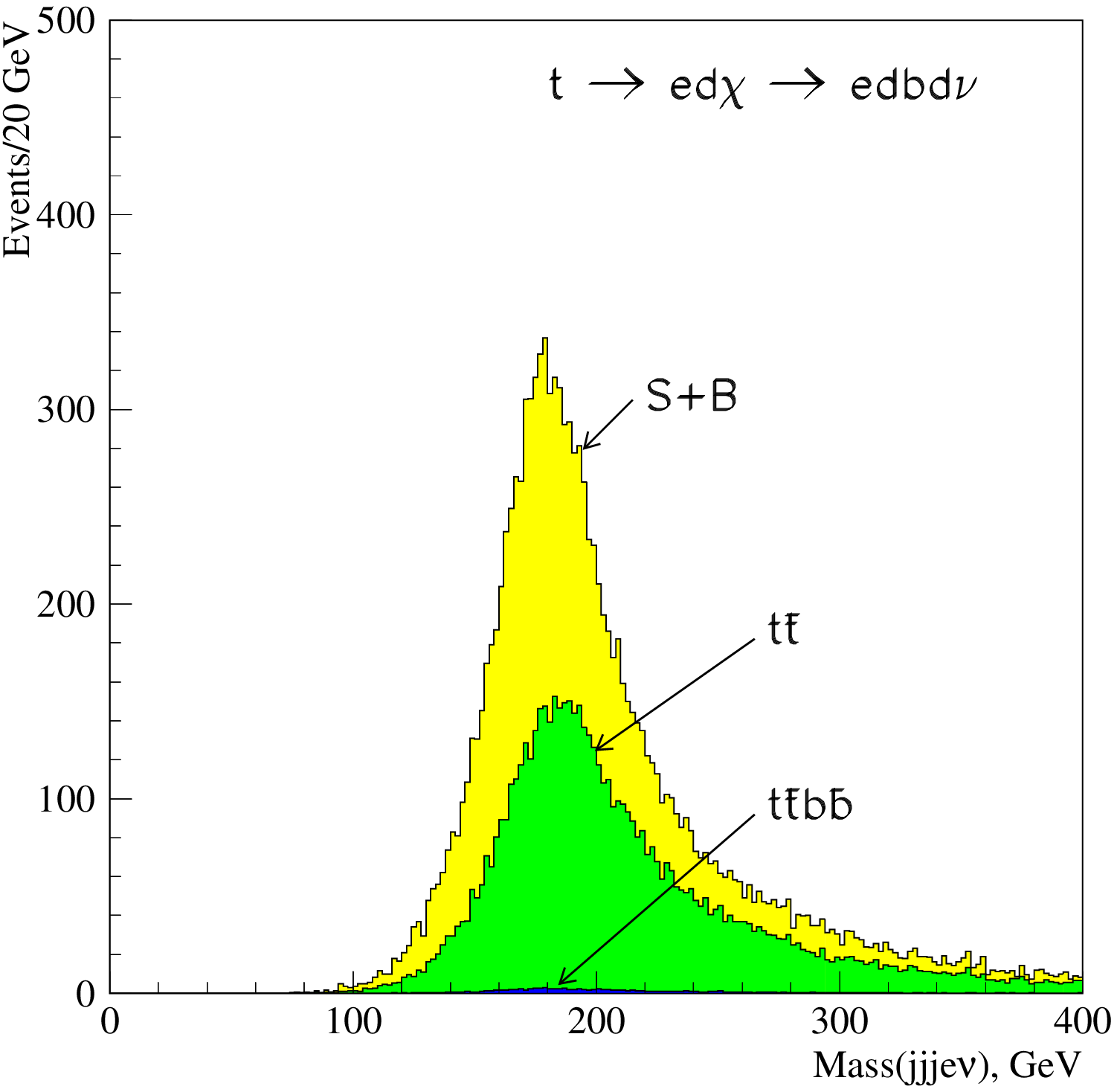,width=0.45\textwidth}
\epsfig{file=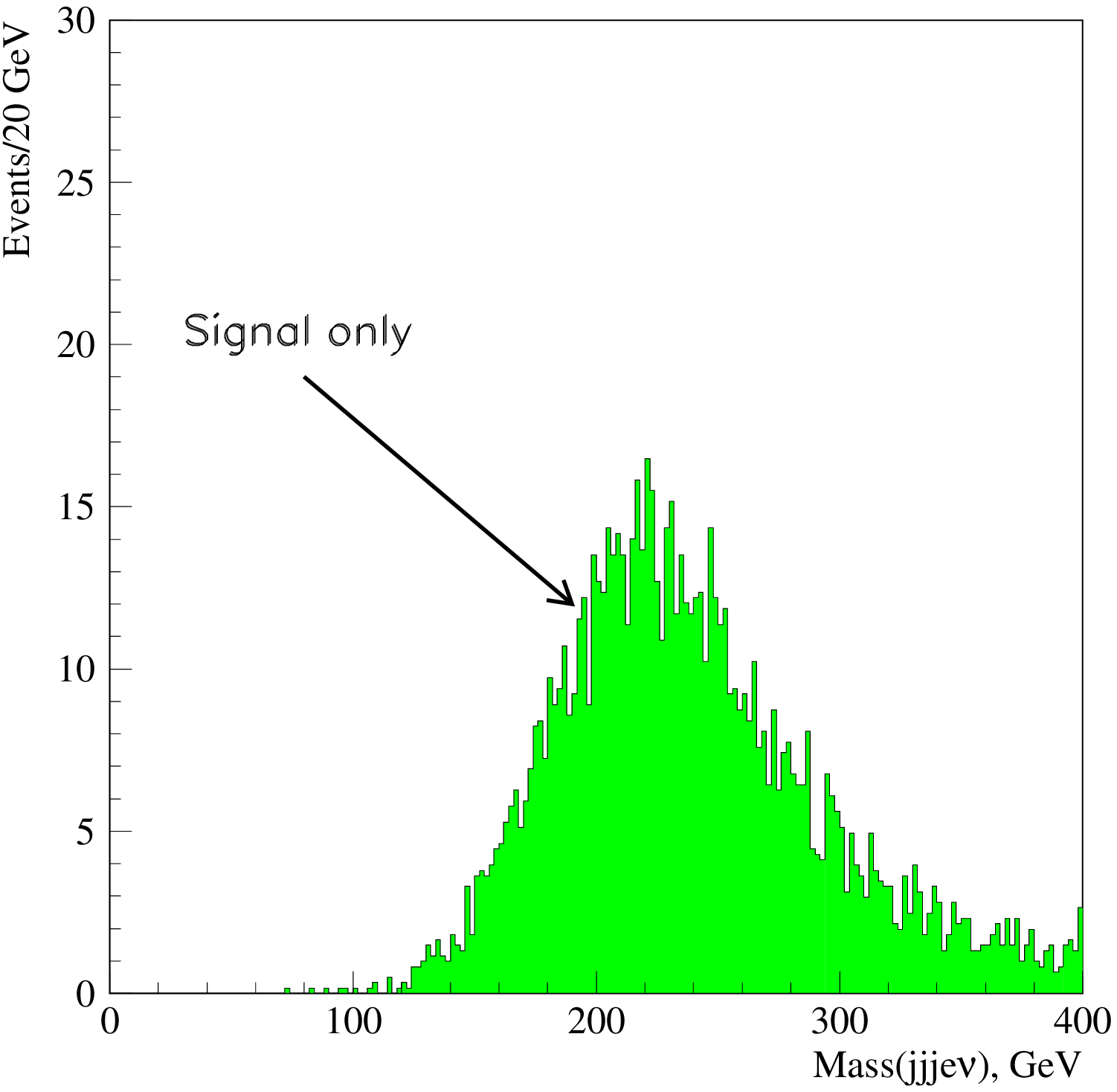,width=0.45\textwidth}
\caption{Detector level invariant mass distribution for the ($jjje\nu$) system 
for $m_{\tilde\ell}=150$~GeV (left) and $m_{\tilde\ell}=200$~GeV (right) cases for 
$\tilde{\chi}^0\to bd\nu$ decay channel.  
Simulations were done for an integrated luminosity of $100 fb^{-1}$ and $300 fb^{-1}$, respectively.
The background of \ttb and \ttbb production for $m_{\tilde\ell}=150$~GeV case is presented.
\label{masstop2}}
\end{figure}

The resulting mass distributions are shown for the ($jjje\nu$) system in 
Fig.~\ref{masstop2}(left) for $\tilde{\chi}^0\to bd\nu$ decay channel 
and the light slepton case($m_{\tilde\ell}=150$~GeV).
The data are presented for an integrated luminosity of 100 fb$^{-1}$.
The distributions for signal and background are shown with separate contributions from 
\ttb and \ttbb events. 
The generation of the \ttbb background has been performed with CompHEP and the cross-section obtained was 5.44 pb.
As one can see below, this background brings contribution of about 2\% to the total 
background for all signal signatures considered.

For the heavy slepton case ($m_{\tilde\ell}=200$~GeV) we present our results  in 
Fig.~\ref{masstop2}(right)for the signal events only for an integrated luminosity of $300 fb^{-1}$.  
Bearing in mind the large $t\bar{t}$ background for $\tilde{\chi}^0\to bd\nu$,
it is clear that no signal can be extracted for the $BDN$ channel and the heavy slepton case. 
We do not show for this case the distributions for backgrounds due to their overwhelming 
dominance.

%%%%%%%%%%%%%%%%%%%%%%%%%%%%%%%%%%%%%%%%%%%%%%%%%%%%%%%%%%%%%%%%%%%% Fig.20
%%%%%%%%%%%%%%%%%%%%%%%%%%%%%%%%%%%%%%%%%%%%%%%%%%%%%%%%%%%%%%%%%%%%

For these events to reconstruct the neutralino, we combine the \emiss with
the two jets with smallest \pT, and for the slepton, we combine the same two jets
with the softest electron and the \emiss. 
The shift from nominal simulated values for the neutralino ( 60 GeV ) and 
the slepton ( 150 GeV ) is due to the difficulties to calculate the energy of neutrino.

%%%%%%%%%%%%%%%%%%%%%%%%%%%%%%%%%%%%%%%%%%%%%%%%%%%%%%%%%%%%%%%%%%%% Fig.21
\begin{figure}[htb] 
\epsfig{file=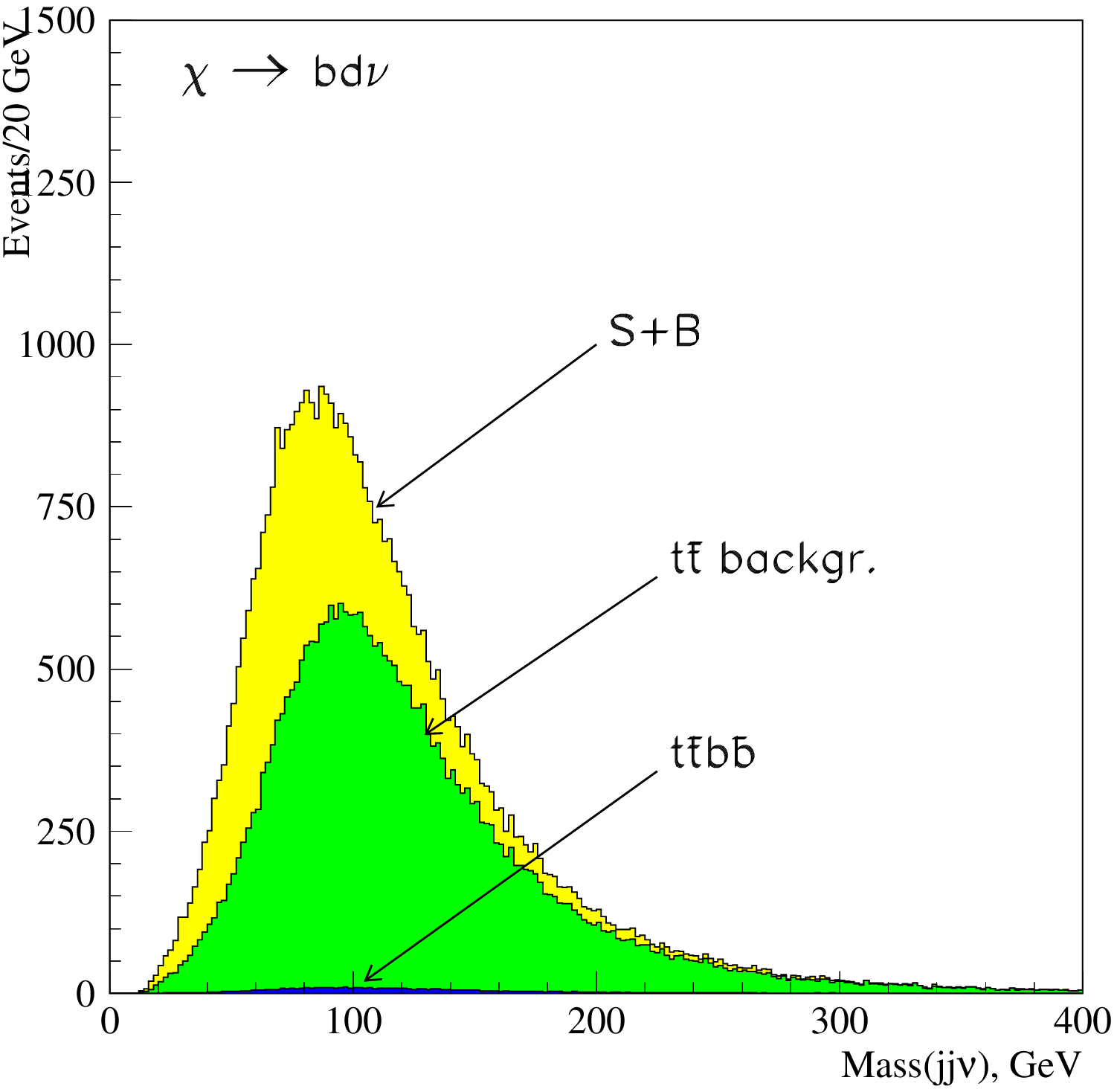,width=0.45\textwidth}
\epsfig{file=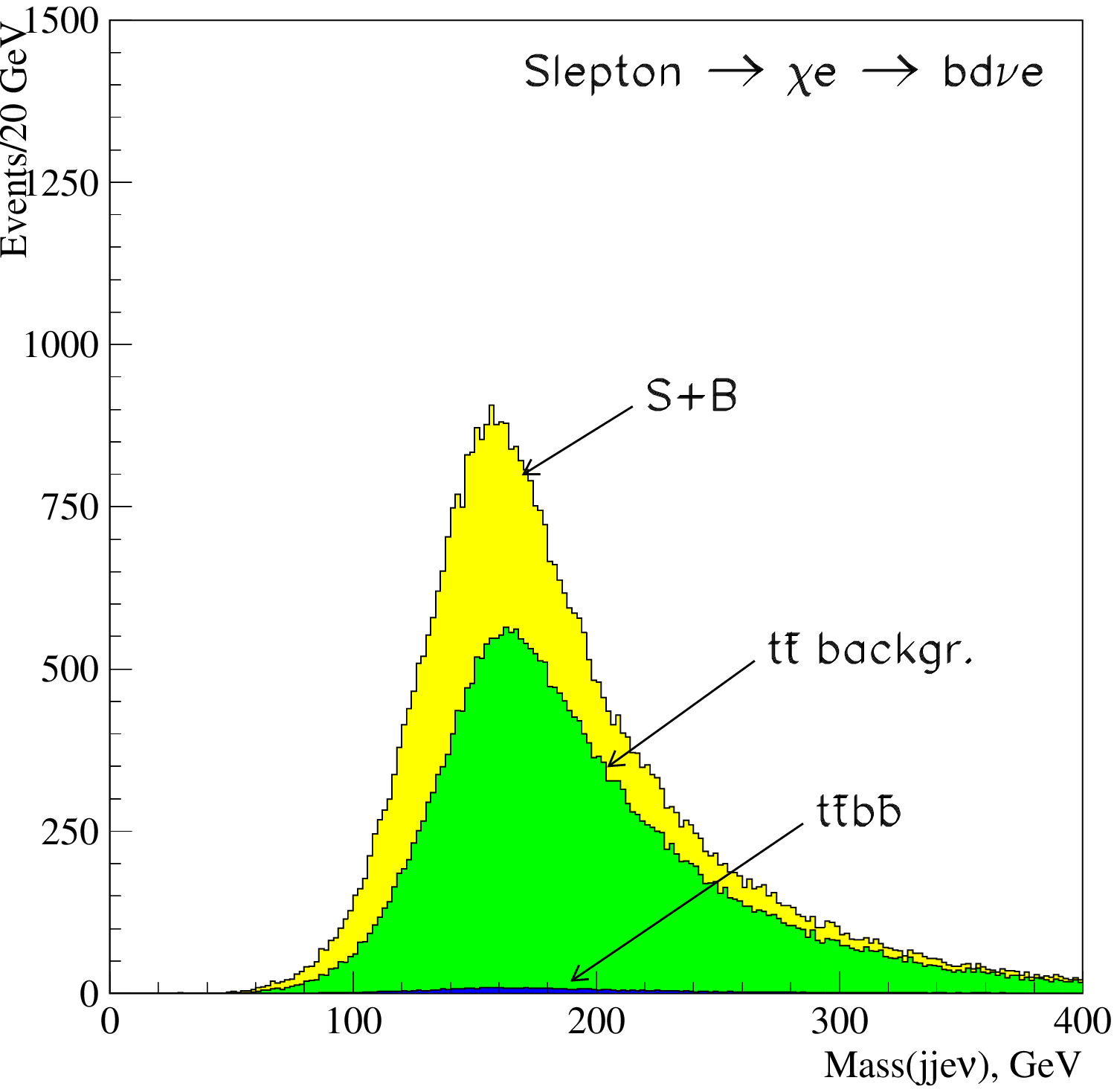,width=0.45\textwidth}
\caption{Detector level invariant mass distribution for the ($jj\nu$) (left) and 
($jje\nu$) (right) systems for the case $m_{\tilde\ell}=150$~GeV and 
$\tilde{\chi}^0 \rightarrow bd\nu$. Background of \ttb events is also shown. 
The distributions are normalized for an integrated luminosity of $100 fb^{-1}$.
\label{massns3}}
\end{figure}
%%%%%%%%%%%%%%%%%%%%%%%%%%%%%%%%%%%%%%%%%%%%%%%%%%%%%%%%%%%%%%%%%%%%

The relative strengths of signal ($\tilde{\chi}^0\to bd\nu$) and the \ttb background
are compared in Figs.~\ref{masstop2}(left) and \ref{massns3}.

For the heavy slepton case we use the same
procedure, but due to a lower production cross-section the signal cannot be seen
due to the \ttb background dominance.

It is worth to mention that the accuracy of the \ttb background measurement for the 
$BDN$ decay channel would be very important. 
An error of $1\%$ on the top quark mass measurements corresponds to an error of $5\%$ 
on the total ttbar cross-section, or vice versa. 
Systematic uncertainties of the measurements of the top quark mass are
dominated by the jet energy scale and FSR effects. Assuming a $1\%$ scale
uncertainty, which is feasible in ATLAS, and varying the jet cone size, should reduce
the total systematic uncertainties\cite{atlas:tdr}.
The precision of the measurements of the total \ttb cross-section is expected also
to be dominated by the knowledge of the absolute scale of the luminosity. 
In any case, the significant amount of data will be required before the \ttb 
background contribution can be estimated accurately.

\subsection{LHC reach}

\begin{center}
\begin{figure}[htb] 
\epsfig{file=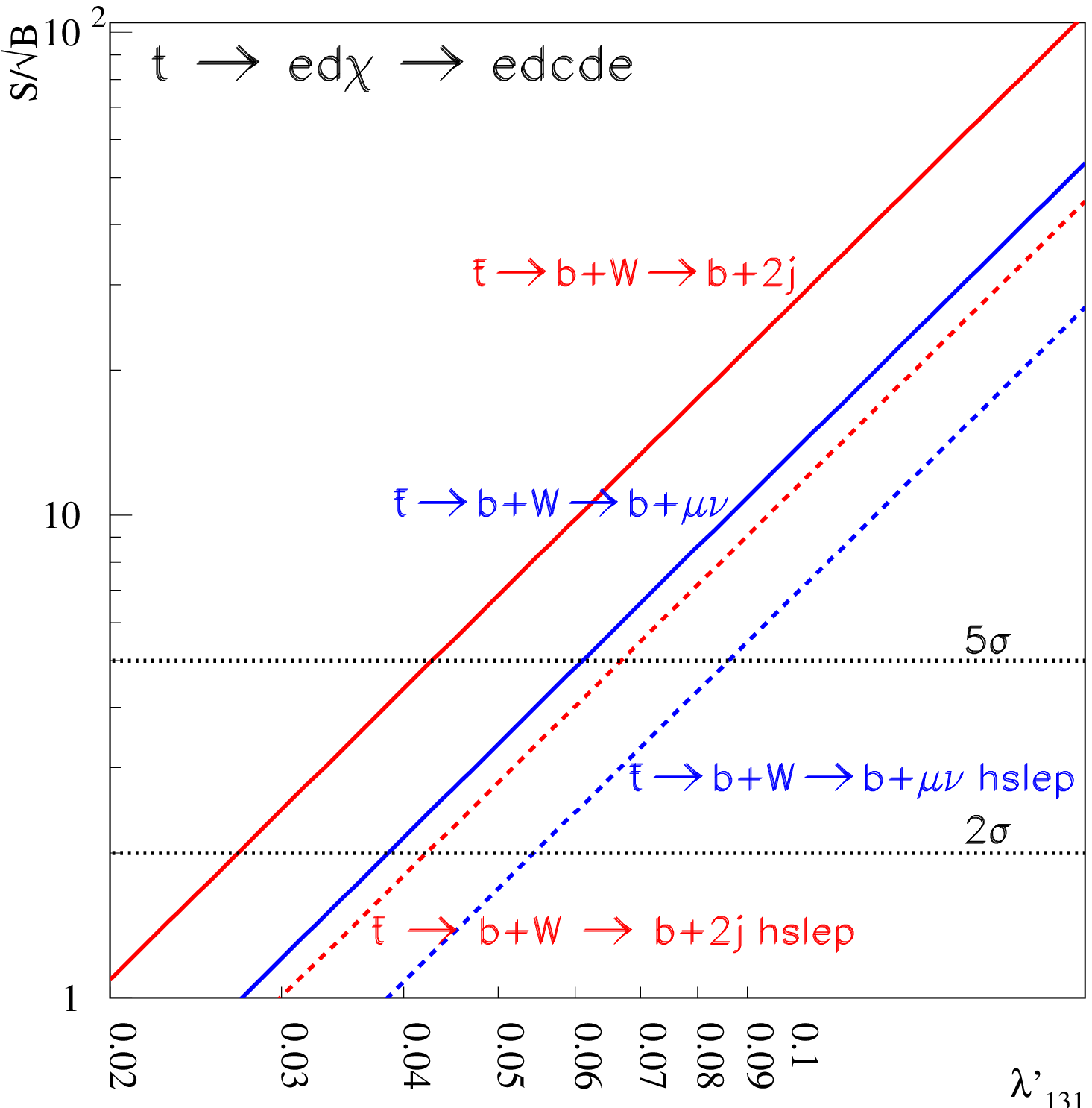,width=0.45\textwidth}
\epsfig{file=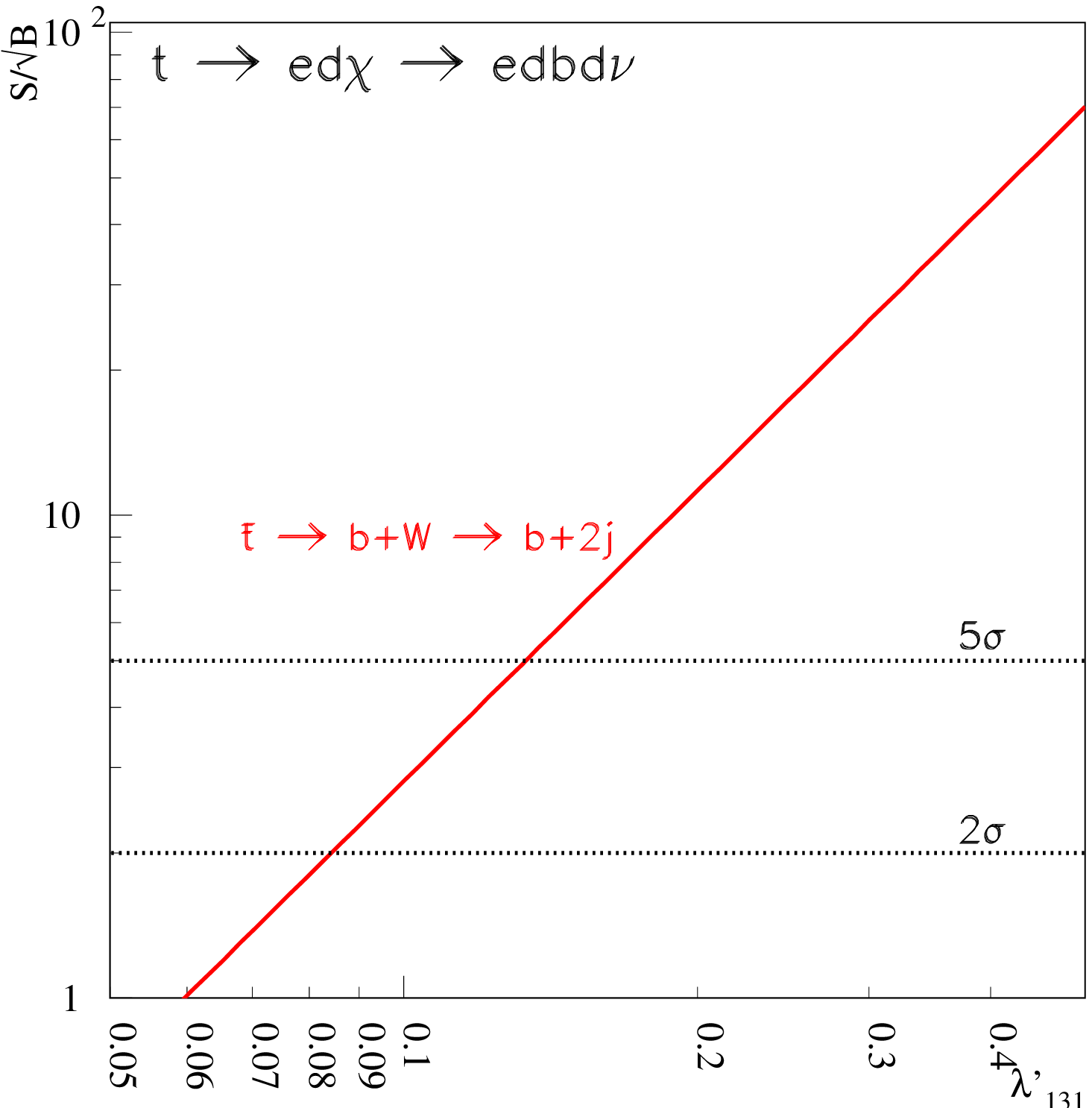,width=0.45\textwidth}
\caption{$S/\sqrt{B}$ vs $\lambda'$ for $CDE$(left) and $BDN$(right) channels.
For $W\to \mu\nu$ decay channel the data are presented for events with
$m_{\tilde\ell}=150$~GeV ( solid line ) and $m_{\tilde\ell}=200$~GeV ( dashed line ).  
\label{lambda}}
\end{figure}
\end{center}

Our final results are presented in Tables~\ref{table1}, \ref{table2} and \ref{table3}
and Fig.~\ref{lambda}.

Tables~\ref{table1} and~\ref{table2} present signal and background cut efficiencies 
($\epsilon^{CDE}$ and $\epsilon^{BDN}$) ) for the channels $\tilde{\chi}^0\to cde$
and $\tilde{\chi}^0\to bd\nu$ channels, respectively, 
for the light slepton case.
The cuts were optimized such that 
$\epsilon^{BDN}_{signal}/\epsilon^{BDN}_{background}\simeq 12$,
while $\epsilon^{CDE}_{signal}/\epsilon^{CDE}_{background}\simeq 7.1 (5.6)$
for leptonic (hadronic) decay channels for $W$-boson.
For the heavy slepton case  the $\epsilon^{CDE}_{signal}/\epsilon^{CDE}_{background}$ 
ratio is similar to the one for the light slepton case.

\begin{table}[ht]
\begin{center}
\caption{
Effective cuts and respective efficiencies ($\epsilon$ ($\%$)) for the signal events 
with $\tilde{\chi}^0\to cde$.}
\begin{tabular}{lccccccc} 
& $\tilde{\chi}^0\to cde$ & $t\to WWj$ & $\tilde{\chi}^0\to cde$  & $t\to WWj$ & \\
&    $W\to \mu\nu$      &     BG     &      $W\to 2jets$      &    BG      & \\ \hline \hline
Number of events        &            &         &          &                   \\   
L = 100 $fb^{-1}$       &   86012    &    67   &  538369  &    417  \\ \hline
Cuts used:     &$\epsilon$($\%$)&$\epsilon$($\%$)&$\epsilon$($\%$)& $\epsilon$($\%$) \\ \hline
2 leptons               &   49.46    &  59.21  &   47.93  &  57.32  \\ \hline
2 leptons + 4 jets      &   36.29    &  21.50  &   17.98  &   8.61  \\ \hline
1 $b-$jet tag           &   19.46    &   9.65  &    9.40  &   3.44  \\ \hline
M$_{slep}$$<$150 GeV    &    3.53    &   0.50  &    1.90  &   0.34  \\ 
\end{tabular}
\label{table1}
\end{center}
\end{table}

\begin{table}
\begin{center}
\caption{Effective cuts and respective efficiencies ($\epsilon$ ($\%$)) for the signal events with $\tilde{\chi}^0\to bd\nu$.}
\begin{tabular}{lccccccc} 
& $\tilde{\chi}^0\to bd\nu$  &  $t\bar{t}$  & $t\bar{t}b\bar{b}$ \\
                            &$W\to2jets$&     BG   &      BG	   \\ \hline \hline
Number of events at         &           &          &		   \\   
L = 100 $fb^{-1}$           &  538576   &  6100000 &    535000     \\ \hline
Cuts used:     &$\epsilon$($\%$)&$\epsilon$($\%$)&$\epsilon$($\%$)& \\ \hline
Cuts used:                  &           &          &	           \\ \hline
1 lepton                    &   61.81   &  45.53   &    48.08      \\ \hline
6 jets + 1 lepton           &   24.22   &  5.34	   &    30.88      \\ \hline
2 b-jet tags                &    7.17   &  2.17	   &    10.72      \\ \hline
M$_{(e\nu b)}<140$ GeV      &    3.57   &  0.53	   &     2.00      \\ \hline
M$_{slep}<$150 GeV          &    1.12   &  0.09	   &     0.30      \\ 
\end{tabular}
\label{table2}
%\end{ruledtabular}
\end{center}
\end{table}

Table~\ref{table3} gives the number 
of expected events, calculated for an integrated luminosity of $L=100 fb^{-1}$,
for the channels $CDE$ and $BDN$ and for the light slepton mass 
$m_{\tilde\ell}=150$~GeV.  
For the heavy slepton case, i.e.  $m_{\tilde\ell}=200$~GeV we only present in 
parentheses the numbers only for the $CDE$ case, since the $BDN$ case is hopeless 
in view of the dominant $t\bar{t}$ background.

\begin{center}
\begin{table}
\caption{Number of expected signal events for $CDE$($\tilde{\chi}^0\to cde$) and $BDN$($\tilde{\chi}^0\to bd\nu$ channels for an integrated luminosity of $L=100 fb^{-1}$. Background(BG) presented are: $tWWq$ events for $\tilde{\chi}^0\to cde$ case and \ttb+\ttbb production for $\tilde{\chi}^0\to bd\nu$ sample of signal events. Heavy slepton case for the $CDE$ channel is given in parenthesis.}
\centering{
\begin{tabular}{cccccccc}
\hline
Signal               &    Mass         &     Signal   &    BG     &     Signal   &    BG     & Signal &  BG  \\ 
                     &   window        &     events   &           &     events   &           & events        \\
                     &    (GeV)        &              &           &              &           &                \\ 
\hline
                     &&$\tilde{\chi}^0\to cde$ &&$\tilde{\chi}^0\to cde$&&$\tilde{\chi}^0\to bd\nu$\\
                     &                 &      $W\to \mu\nu$&       &    $W\to 2j$   &      & $W\to 2j$ &\\
\hline  top quark    &    $175\pm10$   &     949 (-)   & 0.1 (-)   &   2526 (-)   & 0.3 (-)   &  1683 &  1150 \\
\hline               &    $175\pm20$   &    1535 (-)   & 0.1 (-)   &   4389 (-)   & 0.5 (-)   &  2954 &  2138 \\
\hline               &    $175\pm30$   &    1876 (-)   & 0.1 (-)   &   5588 (-)   & 0.7 (-)   &  3835 &  2882 \\
\hline  neutralino   &    $60\pm10$    &    1991 (88)  & 0.2 (0.5) &   6738 (182) & 0.9 (0.9) &  3415 &  2551 \\
\hline               &    $60\pm20$    &    3602 (149) & 0.4 (0.9) &  12433 (294) & 1.8 (1.6) &  6346 &  5172 \\
\hline               &    $60\pm30$    &    4906 (199) & 0.6 (1.3) &  17325 (390) & 2.7 (2.4) &  8558 &  7874 \\
\hline  selectron    &    $150\pm20$   &    3985 (141) & 0.4 (0.7) &  13208 (375) & 1.5 (1.8) &  6253 &  8079 \\
\hline               &    $150\pm30$   &    5330 (194) & 0.5 (1.0) &  18130 (537)& 2.2 (2.7)  &  8503 & 11399 \\
\hline               &    $150\pm40$   &    6278 (240) & 0.7 (1.4) &  21786 (677)& 2.9 (3.5)&   10112 & 13969 \\
\end{tabular}}
\label{table3}
\end{table}
\end{center}

As can be seen
from this table, the $CDE$ signature has negligible background for both 
hadronic and leptonic $W$ decay channels.
The $BDN$ channel, on the contrary, has much larger background coming from SM  
$t\bar{t}$ events. Nevertheless, the $t\bar{t}$ is effectively suppressed
by the fiducial cuts, and the $S/\sqrt{B} \sim 50$ for the case of a light slepton.  

The sensitivity of the LHC to  $\lambda'$ mainly depends on  the slepton
mass, $\mu$,  $M_2$ and $\tan\beta$  which define the top \rpv decay
branching ratio. For our particular choice of these parameters, we estimate
the potential of the LHC to measure $\lambda'$  coupling by studying the 
significance  $S/\sqrt{B}$ versus $\lambda'$ for both cases of neutralino
decay presented in Fig.~\ref{lambda}.

%%%%%%%%%%%%%%%%%%%%%%%%%%%%%%%%%%%%%%%%%%%%%%%%%%%%%%%%%%%%%%%%%%%% Fig.22
%%%%%%%%%%%%%%%%%%%%%%%%%%%%%%%%%%%%%%%%%%%%%%%%%%%%%%%%%%%%%%%%%%%%

One can see the apparent advantage of
the FCNC channel $\tilde{\chi}^0\to cde$ which allows the determination of the exclusion
limits on $\lambda'>0.006$ ( $\lambda'>0.03$ ) at 2$\sigma$~CL or even the possibility to
observe the signal at
5$\sigma$~CL for $\lambda'>0.01$ ( $\lambda'>0.05$ ) for light (heavy) slepton case. 
The case with the SM top$\to bjj$ is more promising
since it provides more statistics. One should remember that we have assumed
the maximal stop-scharm mixing ($V_{\tilde{c}\tilde{t}}$) in our analysis.
Therefore  the limit on $\lambda'$ presented above should be  understood as
the limit on $\lambda'\times V_{\tilde{c}\tilde{t}}$. 

For the $BDN$ channel,  the sensitivity to $\lambda'$ is significantly reduced:
$\lambda'>0.15$ at 2$\sigma$~CL for the light slepton case.
For the heavy slepton case the $BDN$ channel is not observable.
%%% I AM HERE ---> need to cal BR!

The limit on  $\lambda'$ depends on several SUSY parameters.
To reduce the model dependence, it is much more convenient to
present the results in terms of limits on the branching fraction
of \rpv top quark decay -- $BF^t_{\rpv}=BF(t\to q\chi^0\bar{\ell})$.
For the light slepton case, the LHC is sensitive to  $BF^t_{\rpv}\tilde{>} 3\times 10^{-6}$
for the $CDE$ channel and to  $BF^t_{\rpv}\tilde{>}  2\times10^{-3}$
for the $BDN$ channel at 2$\sigma$~CL.

In the case of a heavy slepton scenario, the LHC is sensitive to $BF^t_{\rpv}\tilde{>} 3\times 10^{-8}$.
It looks strange that the LHC is much more sensitive to $BF^t_{\rpv}$
in the heavy slepton case. The puzzle  is hidden in the 
formal definition of $BF^t_{\rpv}=BF(t\to q\chi^0\bar{\ell})$ which assumes
that $BF^t_{\rpv}$ is calculated for the {\it on-mass-shell top-quark}.
However, for the heavy slepton case there are strong off-mass-shell
top-quark effects and, therefore, it is more convenient to define
$BF^t_{\rpv}$ as a ratio of $\rpv$ and SM  {\it cross sections} of the top-quark production 
and decay.
For this definition, light- and heavy-slepton case limits are consistent with each other
and for heavy slepton one has the  $BF^t_{\rpv}$ limit of about $5\times 10^{-6}$  i.e.
a little bit worse than for the light slepton scenario.

It is a quite generic and quasi model independent fact that the  sensitivity of the 
LHC to $BF^t_{\rpv}$ is of the order of $10^{-6}$ 
for $CDE$ channel while for $BDN$ it is about three orders of magnitude worse.

\section{Conclusions}

We have studied the sensitivity of the CERN LHC collider to rare \rpv top quark decay.
We have shown that it is very important to
consider top quark production and its $\rpv$ decay together
(approach (iii) defined in Sect.~2) in order to obtain the 
correct cross section rates and relevant distributions.
The necessity of this treatment is related to the fact
that the $\rpv$ top quark decay involves a heavy intermediate
sfermion which forces the top quark to be off-shell.

Our studies have been done at the  detector fast simulation level.
We have shown that the LHC collider  offers a unique potential to study rare top quark
decays within the framework of supersymmetry
with broken $R$-parity with a sensitivity on the branching fraction $BF^t_{\rpv}$
of the R-parity violating top quark decay in the range of  $\simeq 10^{-6} - 10^{-7}$
for the $CDE$ channel and $\simeq 10^{-3}$ for the $BDN$ channel.
 
We have shown that the $CDE$ channel is not only
very sensitive to the $\lambda'$ coupling, but will also allow the 
measurement (or the setting of a limit ) on the $\tilde{c}-\tilde{t}$ mixing
which is practically unconstrained, at present.
The $BDN$ channel can provide an independent limit on the $\lambda'$ coupling, 
irrespective of the squark-mixing parameter.

\section*{Acknowledgments}  
We would like also to thank G. Azuelos, H.Baer, D.Froidevaux, F. Gianotti, S. Heywood, 
I. Hinchliffe, M.Smizanska, F.Paige, G. Polesello, 
for their comments about the subject. M-H.G., C.L. and R.M. thank 
NSERC/Canada for their support.
This research was supported in part by the U.S. Department of Energy
under the contract number DE-FG02-97ER41022.

\newpage
\newpage
\bibliographystyle{revtex}
\bibliography{top-rpv}
\end{document}